\documentclass[12pt,apj]{emulateapj}

\shorttitle{Radial velocities in very-low mass stars}
\shortauthors{Reiners et al.}

\begin{document}


\title{Detecting Planets Around Very Low Mass Stars with the Radial Velocity Method}


\author{A.~Reiners\altaffilmark{1,3}, J.~L.~Bean\altaffilmark{1,4},
  K.~F.~Huber\altaffilmark{2}, S.~Dreizler\altaffilmark{1},
  A.~Seifahrt\altaffilmark{1}, \& S.~Czesla\altaffilmark{2}}

\altaffiltext{1}{Institut f\"ur Astrophysik, Georg-August-Universit\"at,
  Friedrich-Hund-Platz 1, 37077 G\"ottingen, Germany, Ansgar.Reiners@phys.uni-goettingen.de}

\altaffiltext{2}{Hamburger Sternwarte, Gojenbergsweg 112, 21029 Hamburg, Germany}

\altaffiltext{3}{Emmy Noether Fellow}

\altaffiltext{4}{Marie Curie International Incoming Fellow}




\begin{abstract}
  The detection of planets around very low-mass stars with the radial
  velocity method is hampered by the fact that these stars are very
  faint at optical wavelengths where the most high-precision
  spectrometers operate. We investigate the precision that can be
  achieved in radial velocity measurements of low mass stars in the
  near infrared (nIR) $Y$-, $J$-, and $H$-bands, and we compare it to
  the precision achievable in the optical assuming comparable
  telescope and instrument efficiencies. For early-M stars, radial
  velocity measurements in the nIR offer no or only marginal advantage
  in comparison to optical measurements.  Although they emit more flux
  in the nIR, the richness of spectral features in the optical
  outweighs the flux difference.  We find that nIR measurement can be
  as precise than optical measurements in stars of spectral type
  $\sim$M4, and from there the nIR gains in precision towards cooler
  objects. We studied potential calibration strategies in the nIR
  finding that a stable spectrograph with a ThAr calibration can offer
  enough wavelength stability for m\,s$^{-1}$ precision.  Furthermore,
  we simulate the wavelength-dependent influence of activity (cool
  spots) on radial velocity measurements from optical to nIR
  wavelengths. Our spot simulations reveal that the radial velocity
  jitter does not decrease as dramatically towards longer wavelengths
  as often thought. The jitter strongly depends on the details of the
  spots, i.e., on spot temperature and the spectral appearance of the
  spot. At low temperature contrast ($\sim 200$\,K), the jitter shows
  a decrease towards the nIR up to a factor of ten, but it decreases
  substantially less for larger temperature contrasts.  Forthcoming
  nIR spectrographs will allow the search for planets with a
  particular advantage in mid- and late-M stars.  Activity will remain
  an issue, but simultaneous observations at optical and nIR
  wavelengths can provide strong constraints on spot properties in
  active stars.
\end{abstract}





\section{Introduction}

The search for extrasolar planets with the radial velocity technique
has led to close to 400 discoveries of planets around cool stars of
spectral type F--M\footnote{\texttt{http://www.exoplanet.eu}}.
Fourteen years after the seminal discovery of 51~Peg\,b by
\citet{Mayor95}, the radial velocity technique is still the most
important technique to discover planetary systems, and radial velocity
measurements are required to confirm planetary candidates found by
photometric surveys including the satellite missions CoRoT and
\emph{Kepler}.

The largest number of planets found around solar-type stars are
approximately as massive as Jupiter, and are orbiting their parent
star at around 1\,AU or below. In order to find Earth-mass planets in
orbit around a star, the radial velocity technique either has to
achieve a precision on the order of 0.1\,m\,s$^{-1}$, or one has to
search around less massive stars, which would show a larger effect due
to the gravitational influence of a companion.  Therefore, low-mass M
dwarfs are a natural target for the search for low-mass planets with
the radial velocity technique. In addition, there seems to be no
general argument against the possibility of life on planets that are
in close orbit around an M dwarf \citep[inside the habitable
zone;][]{Tarter07}. So these stars are becoming primary targets for
the search for habitable planets.

So far, only a dozen M dwarfs are known to harbor one or more planets
\citep[e.g.,][]{Marcy98, Udry07}.  The problem with the detection of
radial velocity variations in M dwarfs is that although they make up
more than 70\,\% of the Galaxy including our nearest neighbors, they
are also intrinsically so faint that the required data quality can not
usually be obtained in a reasonable amount time, at least not in the
spectral range most high resolution spectrographs operate at. M dwarfs
have effective temperatures of 4000\,K or less, and they emit the bulk
of their spectral energy at wavelengths redward of 1\,$\mu$m. The flux
emitted by an M5 dwarf at a wavelength of 600\,nm is about a factor of
3.5 lower than the flux emitted at 1000\,nm. Thus, infrared
spectroscopy can be expected to be much more efficient in measuring
radial velocities of low-mass stars.

A second limit on the achievable precision of radial velocity
measurements is the presence of apparent radial velocity variations by
corotating features and temporal variations of the stellar surface.
Such features may influence the line profiles, and that can introduce
a high noise level or be misinterpreted as radial velocity variations
due to the presence of a planet. Flares on active M dwarfs might not
pose a substantial problem to radial velocity measurements
\citep{Reiners09a}, but corotating spots probably do.
\citet{Desort07} modeled the effect of a cool spot for observations of
sun-like stars at optical wavelengths. Their results hint at a
decrease of spot-induced radial velocity signals towards longer
wavelengths.  \citet{Martin06} report the decrease of a radial
velocity signal induced by a starspot on the very active M9 dwarf
LP~944-20 ($v\,\sin{i} \approx 30$\,km\,s$^{-1}$); the amplitude they
find is 3.5\,km\,s$^{-1}$ at optical wavelengths but only an rms
dispersion of 0.36\,km\,s$^{-1}$ at 1.2\,$\mu$m. Thus, observations at
infrared wavelength regions may substantially reduce the effect of
stellar activity on radial velocity measurements, which would allow
the detection of low-mass planets around active stars.

In this paper, we investigate the precision that can be reached in
radial velocity measurements at infrared wavelength regions. The first
goal of our work is to study the detectability of planets around
low-mass stars using infrared spectrographs. We focus on the
wavelength bands $Y, J$, and $H$ because these are the regions where
spectrographs can be built at relatively low cost. Extending the
wavelength coverage to the $K$-band imposes much larger costs because
of severe cooling requirements and large gaps in the spectral format.
We chose to exclude this case from the current paper. Our second
motivation is to see to what extent the radial velocity signal of
active regions can be expected to vanish at infrared wavelength
regions. So far, only rough estimates based on contrast arguments are
available, and no detailed simulation has been performed.

The paper is organized as follows. In \S2, we introduce the spectral
characteristics of M dwarfs and compare model spectra used for our
simulations to observations. In \S3, we calculate radial velocity
precisions that can be achieved at different wavelengths, and we
investigate the influence of calibration methods. In \S4, we simulate
the effect of starspots on radial velocities in the infrared, and \S5
summarizes our results.

\section{Near infrared spectra of M dwarfs}

M dwarfs emit the bulk of their flux at near-infrared (nIR)
wavelengths between 1 and 2\,$\mu$m. However, high-resolution
spectrographs operating in the nIR are not as ubiquitous as their
counterparts in the optical. Therefore, our knowledge about M dwarf
spectra past 1\,$\mu$m is far behind what is known about the visual
wavelength range. Another complication is that strong absorption bands
of water from the Earth's atmosphere cover large fractions of the nIR
wavelength range. Only some discrete portions of the region 1 --
2\,$\mu$m can be used for detailed spectroscopic work.  For our
investigation of radial velocity precision in M dwarfs, we concentrate
on the three bands, $Y$, $J$, and $H$ between the major water
absorption bands.

\begin{figure}
  \plotone{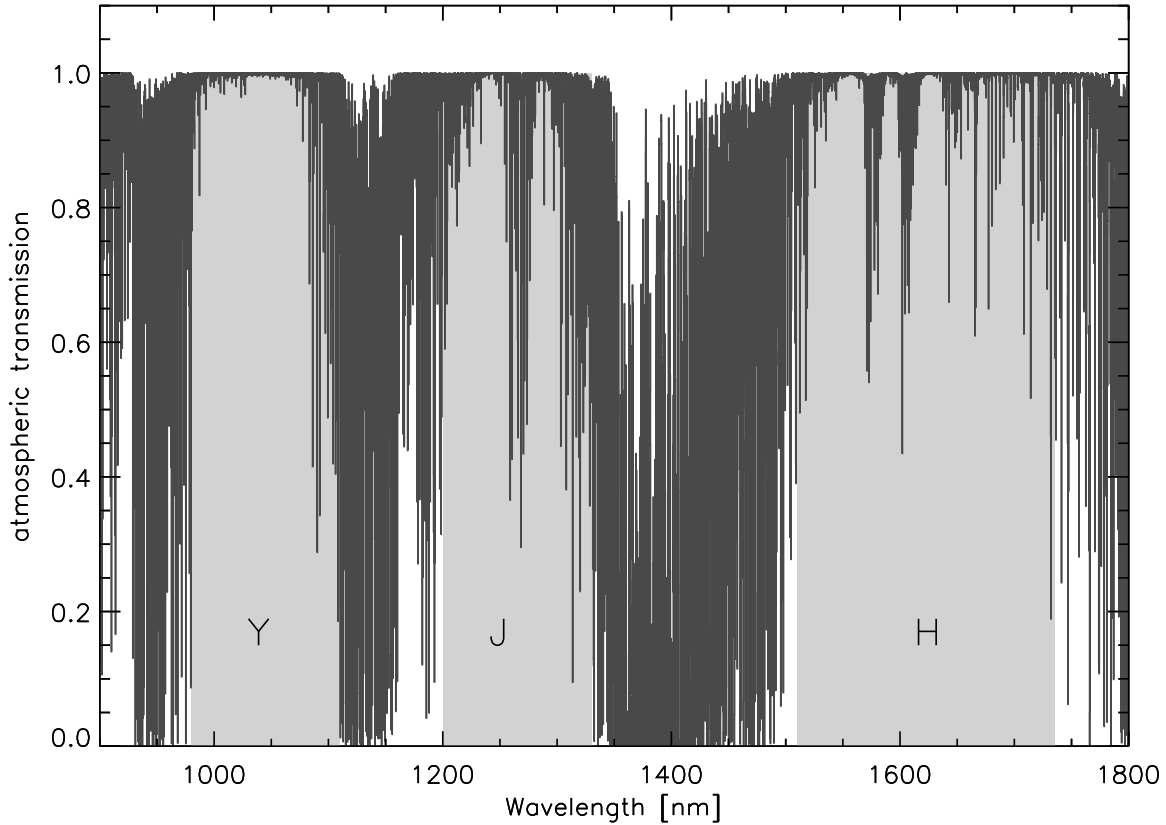}
  \caption{\label{fig:telluric}Telluric absorption spectrum with the
    standard windows in the $Y$-, $J$-, and $H$-bands indicated. }
\end{figure}

We show in Fig.~\ref{fig:telluric} the transmission spectrum of the
Earth's atmosphere together with band identification that we use in
the following. We modeled the telluric features using the FASCODE
algorithm \citep{Clough81,Clough92}, a line-by-line radiative transfer
code for the Earth's atmosphere, and HITRAN \citep{HITRAN2004} as a
database for molecular transitions. Our model is based on a
MIPAS\footnote{\texttt{http://www-atm.physics.ox.ac.uk/RFM/atm/}}
\textsl{nighttime} model atmosphere with updated temperature, pressure
and water vapor profiles for the troposphere and lower stratosphere
based on
GDAS\footnote{\texttt{http://www.arl.noaa.gov/ready/cmet.html}}
meteorological models for Cerro Paranal \citep[see][]{Seifahrt10}.
Fig.~\ref{fig:telluric} shows that the wavelength regions between $Y$,
$J$, and $H$ are not useful for spectroscopic analysis of stars.
Furthermore, it is important to note that the bands themselves are not
entirely free of telluric absorption lines.  While the $Y$-band shows
relatively little telluric contamination, the $J$- and $H$-bands have
regions of significant telluric absorption that must be taken into
account when radial velocities are measured at these wavelengths.  In
our calculations, we mask out the regions that are affected by
significant telluric absorption (see \S\ref{sect:precision.calc}).

Observed low-resolution ($R \approx 2000$) infrared spectra of cool
stars and brown dwarfs were presented by \citet{McLean03} and
\citet{Cushing05}. High-resolution infrared spectra of M dwarfs are
still very rare; \citet{McLean07} and \citet{Zapatero07} show $J$-band
spectra of M, L, and T dwarfs taken with NIRSPEC at $R \approx
20,000$. \citet{Hinkle03} report measurements of rotational velocities
from short portions of $R \approx 50,000$ $K$-band spectra taken with
the Phoenix infrared spectrograph, and \citet{Wallace96} show $K$-band
spectra of cool stars at $R \approx 45,000$. Short $H$-band spectra of
cool stars with a focus on OH lines are presented by \citet{ONeal01},
in particular they show that OH is present in M dwarfs but weaker than
in giants and subgiants.

The high-resolution spectrum that is probably closest to the
appearance of an M dwarf spectrum, and that fully covers the
wavelength range considered in this paper ($Y, J$, and $H$) is the
spectrum of a sunspot. \citet{Wallace92} and \citet{Wallace98}
presented spectra of a sunspot in the visual and nIR regions, and in
the nIR up to 5.1\,$\mu$m, respectively. However, the sunspot spectrum
does not resemble the spectra of M dwarfs at high detail, and it
cannot be used to investigate a range of temperatures among the M
dwarfs. The sunspot spectrum is probably closest to an early-M dwarf
with low gravity \citep{Maltby86}, and we use it below only as a
cross-check for the existence of absorption features predicted in our
models.

To investigate the radial velocity precision that can be reached using
nIR spectra of M dwarfs, we used model spectra calculated with the
PHOENIX code \citep[e.g.,][]{Hauschildt99, Allard01}. This strategy
has the advantage that we can use the full wavelength range without
any restrictions imposed by the unavailability of empirical data, and
that we can model stars of any temperature and surface gravity.  The
caveat, however, is that the model spectra may not adequately resemble
real stellar spectra, in particular at the high spectral resolution we
require.

\begin{figure}
  \plotone{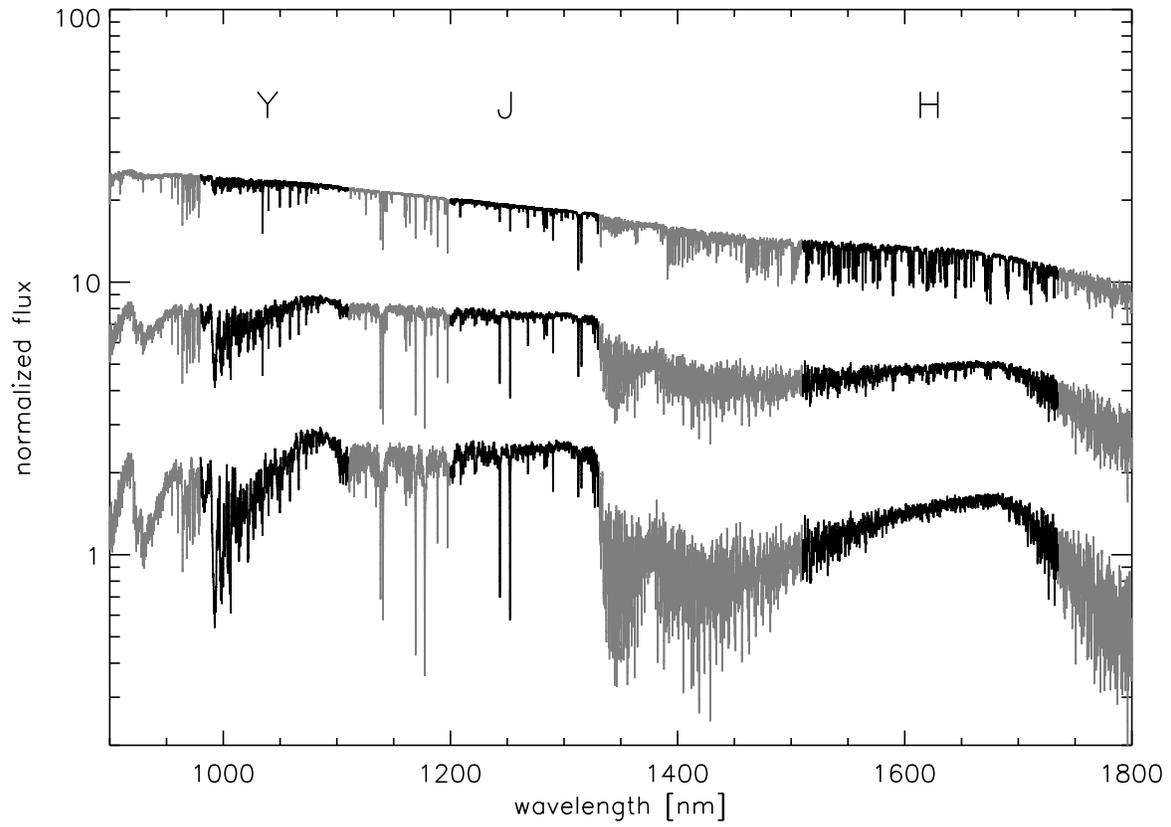}
  \caption{\label{fig:Mstars}M star model spectra for three different
    effective temperatures: 3500\,K (M3, upper panel), 2800\,K (M6,
    middle panel), and 2400\,K (M9, lower panel). The black regions
    show the photometric windows $Y$, $J$, and $H$. Gray regions are
    the telluric gaps separating these windows.}
\end{figure}

\begin{figure}
  \plotone{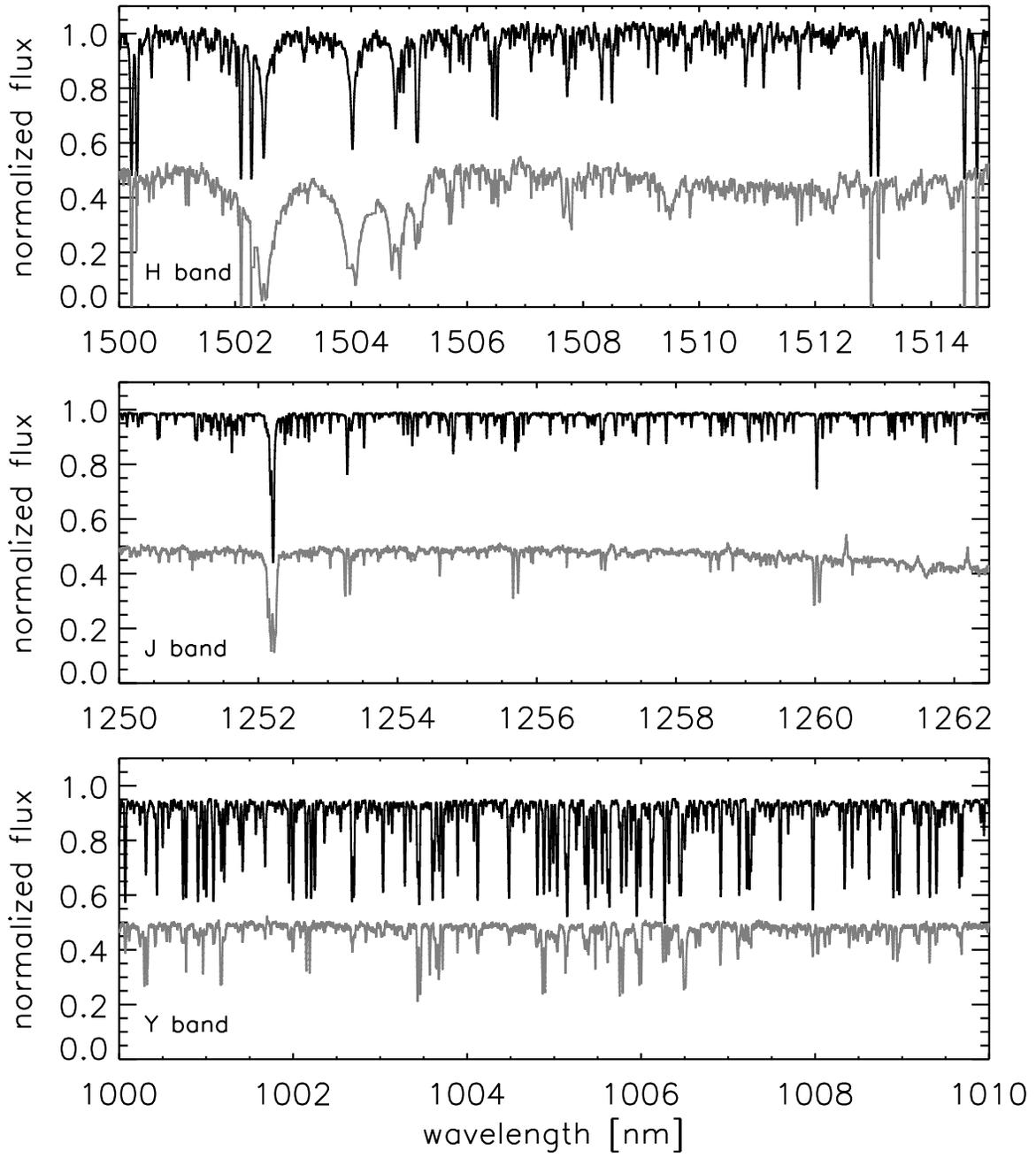}
  \caption{\label{fig:YJH}M star model spectrum for 3500\,K (M3, black
    line) and a sunspot spectrum (grey line) in the $Y$-, $J$-, and
    $H$-bands.}
\end{figure}

We show in Fig.~\ref{fig:Mstars} three model spectra of cool dwarfs,
spectral types are approximately M3, M6, and M9, i.e., effective
temperatures of T = 3500, 2800, and 2600\,K, respectively. We use
models with a surface gravity of $\log{g} = 4.75$ throughout this
paper. The spectra reproduce the data shown by \citet{McLean03} and
\citet{Cushing05} reasonably well when degraded to low-resolution.  In
Fig.~\ref{fig:YJH}, we show short parts of the model spectra at $Y$-,
$J$-, and $H$-bands at higher detail for the case of an M3 star (T =
3500\,K).  The three spectral windows in Fig.~\ref{fig:YJH} all cover
the same number of resolution elements assuming the spectral
resolution is the same in all bands.  Obviously, the $Y$-band is very
rich in structure.  There are only a few deep lines in the $J$-band,
and the number of sharp features in the $H$-band is lower than in the
$Y$-band, but higher than in the $J$-band.

The model shown in Fig.~\ref{fig:YJH} is at the hot end of targets we
are interested in.  Comparison to Fig.~\ref{fig:Mstars} indicates that
with lower temperature, the features in the $Y$-band become stronger.
The same is true for features in the $J$-band, but the $H$-band
becomes relatively featureless at late-M spectral class, which is
mainly due to the disappearance of OH lines. In Fig.~\ref{fig:YJH}, we
also show the sunspot spectrum in comparison to the M3 model spectrum.
The FeH lines in the sunspot $Y$-band are somewhat weaker than in M
dwarfs and are strongly affected by Zeeman broadening
\citep{Reiners06}, the latter leading to significantly reduced depths
for many lines in the sunspot spectrum. In the $J$- and the $H$-bands,
the main features and their depth are relatively well described, so
that our model spectra likely reproduce the stellar spectrum
relatively well, at least at this temperature.

\begin{figure}
  \plotone{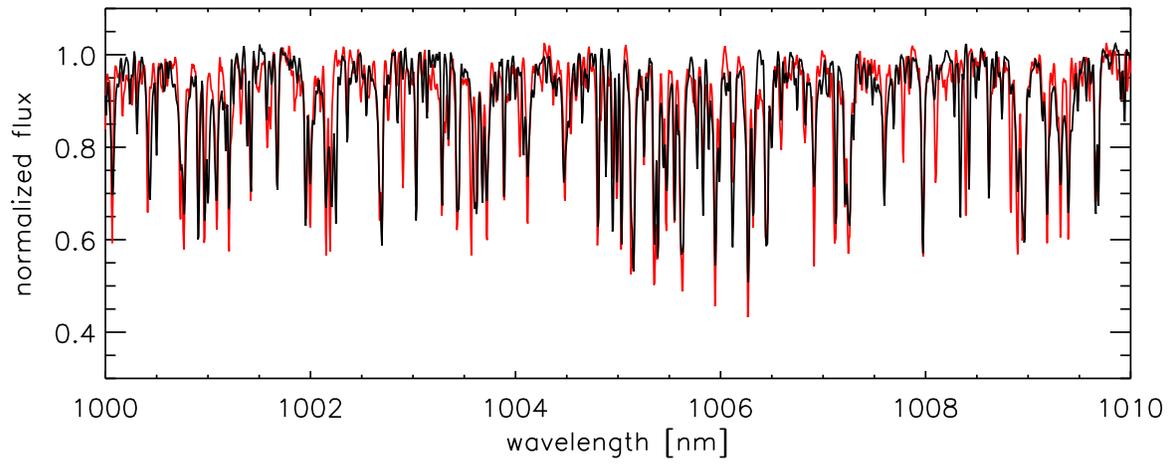}
  \caption{\label{fig:CRIRES}CRIRES spectrum of GJ 1002, M5.5 (red),
    and a model spectrum for T=3200K and logg=4.75 (black). }
\end{figure}

Another comparison of our models to an observed high-resolution
spectrum of a mid-M dwarf is shown in Fig.~\ref{fig:CRIRES}, where we
compare an observed spectrum of GJ~1002 (M5.5) to a model spectrum at
a temperature of 3200\,K. The observed spectrum was obtained by us
using CRIRES \citep{Kaufl06} at the Very Large Telescope (VLT) and
reduced following standard infrared reduction procedures including
bias- and sky-subtraction, flatfielding, and wavelength calibration
using ThAr lines. The model lines (predominantly from FeH) provide a
good fit to the observed spectrum of GJ~1002. Thus, we feel confident
that our model spectra reproduce the FeH band in the $Y$-band in M
dwarfs reasonably well. More generally, all these comparisons suggest
that the PHOENIX model spectra are accurate enough for simulations of
radial velocities measured from the nIR spectra of M dwarfs.

\section{Radial velocity precision}
\label{sect:precision}

\subsection{Calculation of radial velocity precision}
\label{sect:precision.calc}

The achievable precision of a radial velocity measurement from a
spectrum of given quality was calculated by \citet{Connes85} and
\citet{Butler96}. This value is a general limit for radial velocity
measurements; it reflects the limited intrinsic information content of
any observed spectrum.  For example, if a star exhibits more lines in
a certain wavelength range, this will lead to a higher precision
compared to a star with fewer lines. Similarly, a set of shallow,
perhaps blended, lines will put a weaker constraint on the radial
velocity than a set of narrow, deep lines. In their Eq.~6,
\citet{Butler96} provide a formula for the radial velocity uncertainty
as a function of intrinsic content, instrument resolution ($R$), and
signal-to-noise ratio ($S/N$).  This value is inversely proportional to
the weighted sum of the spectrum derivative, which means that the
precision is higher as the spectrum has more deep and sharp features.
In the following, we first calculate the intrinsic radial velocity
precision achievable in a stellar spectrum at given $R$ and $S/N$. In a
second step, we ask how this precision is affected by the limited
precision of the wavelength calibration (\S\ref{sect:calibration}).

In order to compare the potential of different wavelength bands, we
take a model spectrum, assume a $S/N$ at one given wavelength, and
calculate the $S/N$ at other wavelengths according to the spectral flux
distribution. We assume constant instrument efficiency at all
wavelengths, and let the signal quality vary according to the stellar
flux distribution. We also assume constant spectral resolution and
sampling at different wavelength ranges. Calculating the $S/N$ from the
spectral flux distribution, it is important to use the number of
photons per spectral bin instead of energy per spectral bin, which is
provided by the PHOENIX model spectra. To convert energy flux into
photon flux, we need to divide by the energy carried per photon, which
is $hc/\lambda$. Neglecting the constants, this means we need to
multiply the spectrum with $\lambda$.

For our calculations, we assume an average $S/N$ of 100 at a resolving
power of $R = 60,000$ in the $Y$-band. To account for the spectral
resolution, we apply an artificial broadening to the spectra so that
the lines become wider, and we assume that the instrument collects a
constant number of photons per wavelength interval, i.e., more photons
are available per bin at lower spectral resolution. Note that this
approach ignores the possibly higher losses from using a narrower slit
or other design differences to achieve higher resolution that are
likely in practice. That is, a real spectrograph would likely deliver
more photons per wavelength interval when used in a lower resolution
mode. As this effect is likely a lower-order consideration than the
effect of varying dispersion, and is also difficult to predict in a
general sense, we do not consider it here.

\subsection{Impact of telluric lines}

Near infrared wavelength regions are more severely affected by
telluric contamination than the optical range. The limiting effect of
telluric contamination lies in the difficulty removing it to a level
at which it does not affect the radial velocity measurement on a
m\,s$^{-1}$ level. For example, \citet{Bean09} report a limit of
5\,m\,s$^{-1}$ that can be reached when spectral regions with telluric
lines are included in the analysis. To reach higher precision,
contaminated regions need to be excluded from the analysis.

In our calculations, we mask the regions affected by telluric lines
and do not use them for our analysis of radial velocity precision. We
chose to ignore all spectral regions that fall in the vicinity of
$\pm$30\,km\,s$^{-1}$ around a telluric absorption of 2\,\% or more,
which is approximately the span introduced by maximum barycentric
velocity differences. The telluric transmission spectrum was
artificially broadened to a spectral resolving power of $R = 100,000$
before the 2\,\% cutoff was applied. The exact regions that fulfill
this criterion also depend on the atmospheric conditions. The
wavelength bands together with the fractions ignored due to
contamination with telluric lines are summarized in
Table\,\ref{tab:bands}. The telluric contamination in the $V$-band is
rather small (2\,\%), and does not badly affect the $Y$-band either
($< 20$\,\%). On the other hand, roughly half of the spectrum in the
$J$- and $H$-bands (55\,\% and 46\,\%, respectively) is affected by
telluric contamination. The effect on the theoretical information
content of the stellar spectrum hence is $\sim \sqrt{2}$, which is
still not an order-of-magnitude effect but can no longer be neglected.
At these wavelengths, one has to decide whether the RV precision is
higher after discarding the contaminated spectral range, or if one
should attempt to correct for the telluric contamination.

We note that the same excercise for the $K$-band (2050--2400\,nm)
shows that about 80\,\% of this wavelength range is affected by
significant telluric contamination. Clearly, radial velocity work in
the $K$-band faces very severe limitations due to telluric lines.

\begin{deluxetable}{ccccc}
  \tablecaption{\label{tab:bands}Wavelength coverage of the spectral
    windows used in this work and the fraction of the wavelength range
    affected by telluric contamination.}
  \tablewidth{0pt}
  \tablehead{\colhead{Band} & $V$ & $Y$ & $J$ & $H$}
  \startdata
  $\lambda$-range [nm] & 505--595 & 980--1110 & 1200--1330 & 1510--1735 \\
  telluric loss & 2\,\% & 19\,\% & 55\,\% & 46\,\% 
  \enddata
\end{deluxetable}

\subsection{Wavelength calibration methods}
\label{sect:calibration}

A critical part of radial velocity measurements is the wavelength
calibration. When considering such measurements in new spectral
regimes the influence of available calibration precision must be
considered in addition to the intrinsic information content of the
spectra of the stars of interest. There are generally two types of
wavelength standards that are being used very successfully at optical
wavelengths for high-precision radial velocity work; 1) the
calibration signal is injected in the spectrograph following a
separate path, e.g., using a ThAr emission lamp \citep{Pepe02}, and 2)
the calibration signal is imposed on the stellar light using a gas
absorption cell \citep[e.g. I$_{2}$,][]{Butler96}. At nIR wavelengths,
both techniques so far have not been used at the same level of
efficiency as in the optical, mainly because no instruments are yet
available that can provide comparable spectral resolving power and
wavelength range as instruments operating at optical wavelengths.
However, such spectrographs are foreseen for the future, and we
estimate the precision of both techniques and their current
applicability in the nIR. We cast these calculations in terms of
equivalent radial velocity precision so that they may be compared
directly to the estimated information content of the stellar spectra.
For our purpose, we have not considered calibration using a laser comb
\citep{Steinmetz08} or an etalon, which essentially follow the same
principle as the ThAr calibration. A laser comb or etalon that cover
the full desired wavelength range would largely solve the problems of
inadequate wavelength coverage. Unfortunately, both are not yet
available and we restrict the discussion to the ThAr and gas cell
options.

\subsubsection{ThAr lamp}

In the nIR, the ThAr method could in principle just be copied from the
optical regime. Standard ThAr lamps produce fewer lines in the nIR,
but that does not necessarily mean that the precision over large
wavelength regions must be lower than at optical wavelengths. For
example, \cite{Kerber08} provide a list of ThAr lines that includes
more than 2400 lines in the range 900 -- 4500\,nm. \citet{Lovis06}
found that the Ar lines produced by a ThAr lamp are unsuitable for
high precision wavelength calibration because they show high intrinsic
variability on the order of tens of m\,s$^{-1}$ between different
lamps. Nevertheless, \citet{Kerber08} show that in the wavelength
range we consider here, the fraction of Ar lines in the ThAr lamp
spectrum is only on the order of $\sim 15$\,\%, and these authors also
discuss that the pressure sensitivity only appears in the
high-excitation lines of \ion{Ar}{1}. So although there are fewer
lines than in the optical, the still high number of possible lines
suggests that a ThAr lamp should be evaluated as a possible wavelength
calibration.

To estimate the calibration precision that can be reached with a ThAr
spectrum in a given wavelength interval, we count the number of lines
contained in this interval in the list of \cite{Kerber08}, and we
estimate the uncertainty in determining the position of every line
(converted to radial velocity units) according to its intensity. We
assume that we only take one exposure of the ThAr lamp, and we scale
the line intensities to a given dynamic range. The range ultimately
used for our calculations was selected to achieve a high number of
useful lines while losing as few lines as possible due to detector
saturation. We quadratically add the uncertainties of all lines to
calculate the total uncertainty of a ThAr calibration at the chosen
spectral region.  ``Saturated'' lines are not taken into account, but
we note that infrared detectors offer advantages over CCDs in this
regard. One advantage of infrared detectors is that saturated pixels
do not bleed into neighboring pixels like in CCDs. Therefore, although
a particular calibration line may saturate some pixels, the problem
would be localized only to the pixels the line falls on and the
signals recorded for lines falling on neighboring pixels would be
unaffected. In practice then, some lines may be allowed to saturate
and thus be ignored during the wavelength calibration if a higher
overall signal would result in a net gain of useful lines. A second
issue is that individual pixels in infrared detectors can be read out
at different rates. Therefore, there is the possibility of an
increased dynamic range for a given exposure level. However, it is
unclear what would be the influence of the differences in response and
noise properties that pixels read out at different rates would have.
So for this work we take the conservative approach that lines which
would apparently saturate given our selected dynamic range can not be
used for the wavelength calibration.

Our estimation of the utility of a ThAr lamp for wavelength
calibration in order to obtain high-precision radial velocities is
based on the assumption that the calibration is only needed to track
minor changes in the instrument response. Such is true for an isolated
and stabilized instrument with a nearly constant pupil illumination
\cite[e.g. like HARPS,][]{Mayor03}. The main reason is that the light
from a ThAr lamp will not pass directly through the instrument in the
exact same way and/or at the exact same time as the light from a star.
Therefore, the utility of ThAr as a calibration for radial velocity
measurements will be reduced from that discussed here for instruments
that experience significant temporal variations.

\subsubsection{Gas absorption cell}
 
The gas absorption technique requires a gas that provides a large
number of sharp spectral lines. Currently, no gas has been identified
that produces lines in the full nIR wavelength range at a density
comparable to I$_2$ spectrum in the optical (I$_2$ only provides lines
in the optical wavelength range), although there have been some
investigations into gases suitable for small windows in this region.
\citet{DAmato08} report on a gas absorption cell using
halogen-hydrates, HCl, HBr, and HI, that has absorption lines between
1 and 2.4\,$\mu$m, but these gases only produce very few lines so that
a calibration of the wavelength solution and the instrumental profile
can only be done over a small fraction of the spectrum.
\citet{Mahadevan09} discuss various options for nIR gas cells and
conclude that the gases H$^{13}$C$^{14}$N, $^{12}$C$_{2}$H$_{2}$,
$^{12}$CO, and $^{13}$CO together could provide useful calibration in
the $H$-band. Another gas that provides some utility in the nIR is
ammonia (NH$_3$), which exhibits a dense forest of spectral lines in
the $K$-band. We are currently using an ammonia cell for a radial
velocity planet search with CRIRES at the VLT. More details about the
cell and these radial velocity measurements are contained in another
paper \citep{Bean09}.

For an estimate of the calibration precision that could potentially be
achieved over a broad wavelength range using a gas cell, we assume
that a gas or combination of gases might be found with absorption
lines similar to ammonia, but throughout the entire nIR region. We
calculate the radial velocity precision from a section of an ammonia
cell spectrum with various $S/N$ and R just as for the stellar spectra
\citep[i.e. using Eq.~6 in][]{Butler96}. The basis for this
calculation is a 50\,nm section of a spectrum of our CRIRES ammonia
cell (18\,cm length and filled with 50\,mb of ammonia) measured with
an FTS at extremely high-resolution ($R \sim 700,000$). Convolved to a
resolving power of $R = 100,000$ and $S/N$ = 100, the calculated
precision is 9\,m\,s$^{-1}$. We note that this value would change if a
longer cell or higher gas pressure would be used, but this change
would be relatively small for conceivable cells. To extrapolate the
precision estimate to arbitrary wavelength regions we scale the
calculated value by the corresponding size of the regions.  For
example, the uncertainty from a region of 100\,nm would be a factor of
$\sqrt{2}$ less than that calculated for the 50\,nm region.

We emphasize that our estimates on the performance of a gas cell in
the nIR are purely hypothetical.  Currently, in the wavelength
range under consideration, we know of no real gas that shows as dense
a forest of lines in the $Y$-, $J$-, and $H$-bands as ammonia does in the
$K$-band.

\subsection{Radial velocity precision in low-mass stars}

\begin{deluxetable}{rrccccrcccc}
  \tablecaption{\label{tab:accuracies}Wavelength-dependent $S/N$ and
    radial velocity precision that can be achieved from data of this
    quality. The upper part of the table shows the results for an M3
    star, the middle for an M6, and the lower part for an M9 star. }
  \tablewidth{0pt} \tablehead{\colhead{Resolution} &&
    \multicolumn{4}{c}{$S/N$} && \multicolumn{4}{c}{RV precision [m
      s$^{-1}$]}} \startdata
    && $V$ & $Y$ & $J$& $H$ && $V$ & $Y$ & $J$& $H$\\
  \cutinhead{Spectral Type M3}
 60000 && 50 & 100 & 101 &  95 &&  3.6 &  5.7 & 22.9 & 10.0\\
 80000 && 43 &  86 &  87 &  82 &&  2.9 &  4.4 & 18.1 &  8.4\\
100000 && 39 &  77 &  78 &  74 &&  2.5 &  3.8 & 15.5 &  7.6\\
  \cutinhead{Spectral Type M6}
 60000 && 20 & 100 & 114 & 107 &&  4.7 &  3.8 & 11.2 &  9.7\\
 80000 && 18 &  86 &  99 &  93 &&  3.7 &  3.0 &  8.8 &  7.8\\
100000 && 16 &  77 &  88 &  83 &&  3.2 &  2.6 &  7.5 &  6.9\\
  \cutinhead{Spectral Type M9}
 60000 && 12 & 100 & 134 & 128 &&  8.0 &  2.2 &  4.6 &  4.0\\
 80000 && 10 &  86 & 116 & 111 &&  6.2 &  1.7 &  3.5 &  3.5\\
100000 &&  9 &  77 & 104 &  99 &&  5.3 &  1.5 &  2.9 &  3.3
  \enddata
\end{deluxetable}

\begin{figure}
  \includegraphics[width=.5\textwidth,bbllx=50,bblly=70,bburx=615,bbury=434,clip=]{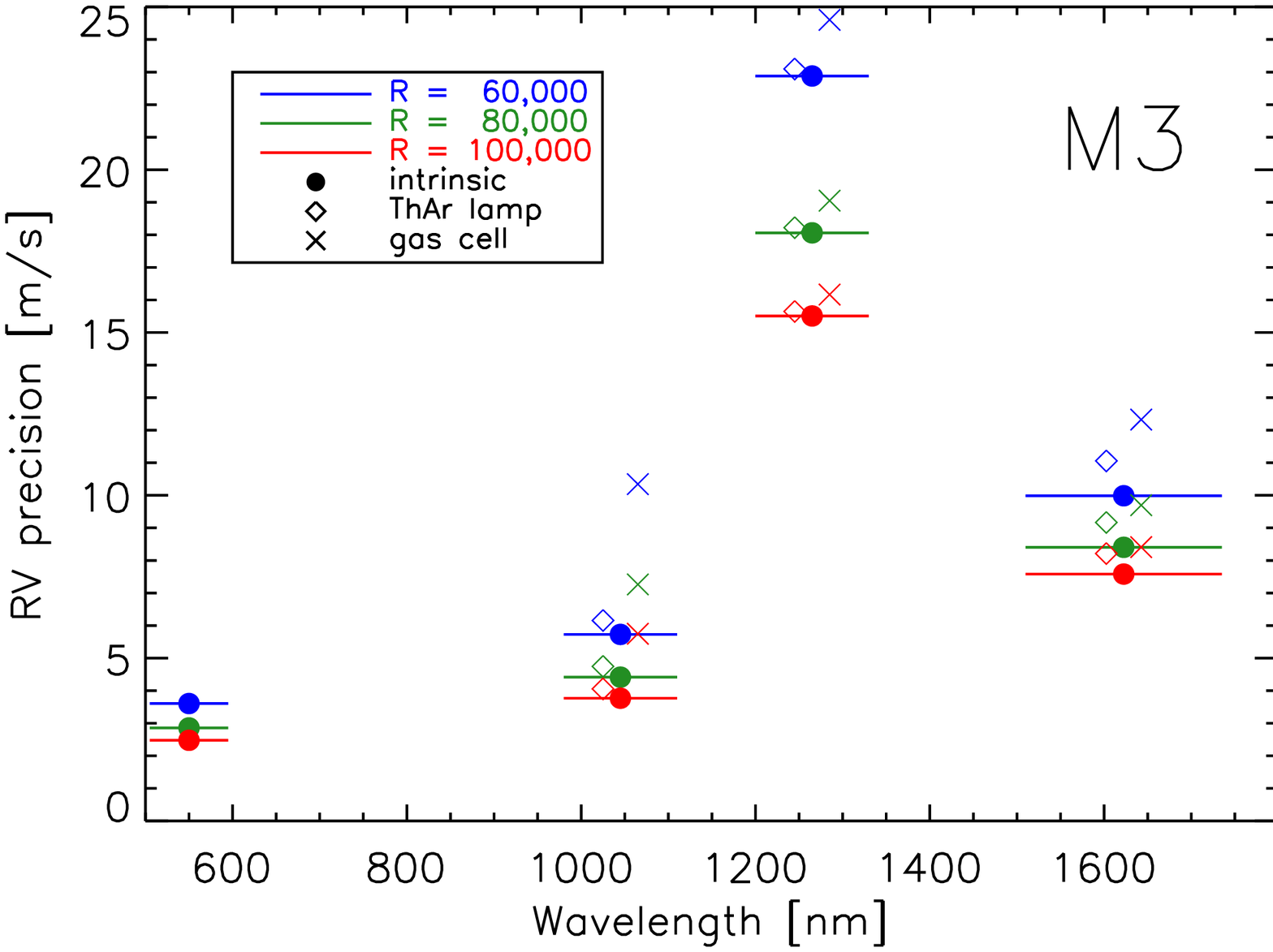}\\
  \includegraphics[width=.5\textwidth,bbllx=50,bblly=70,bburx=615,bbury=434,clip=]{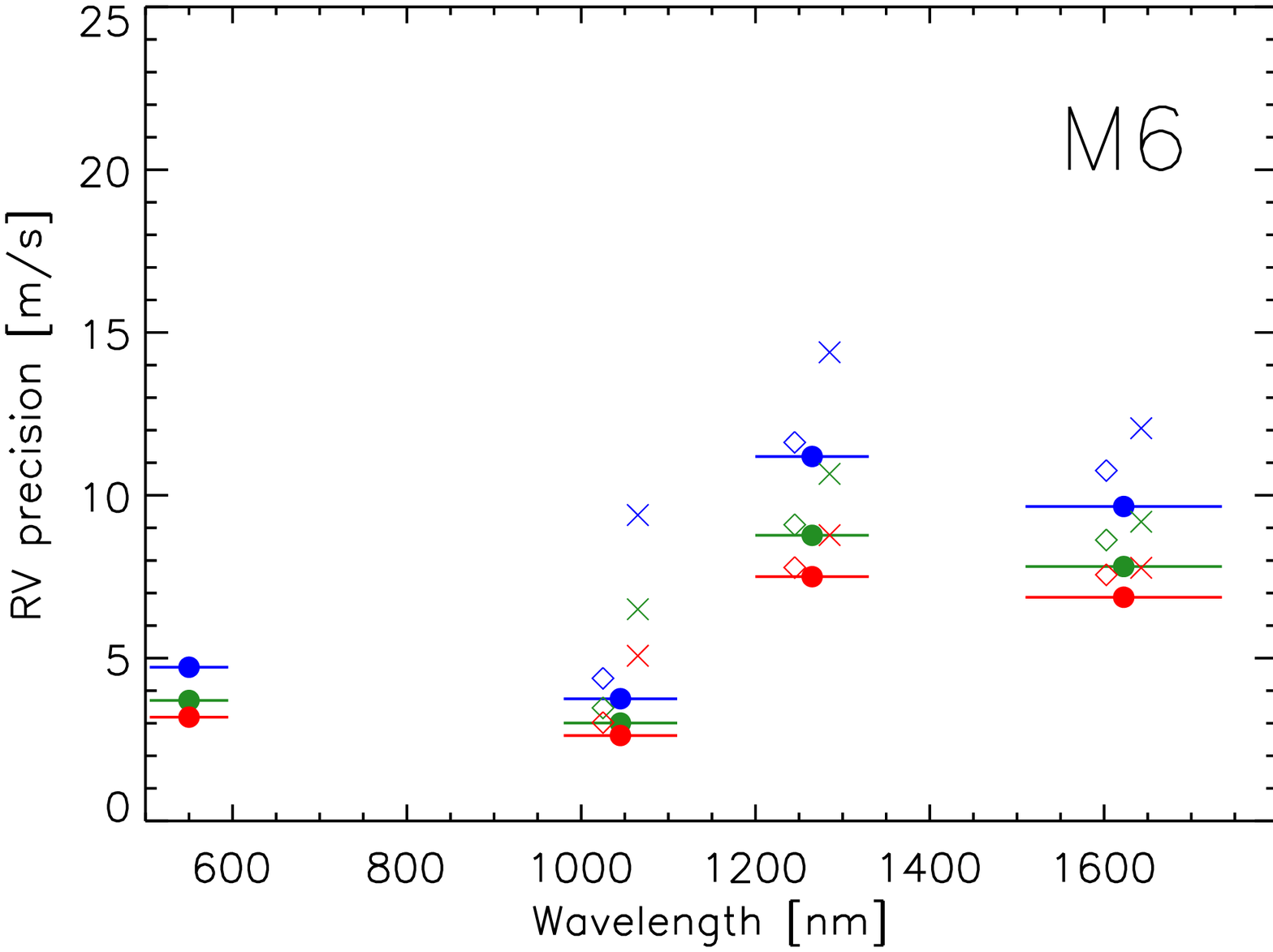}\\
  \includegraphics[width=.5\textwidth,bbllx=50,bblly=15,bburx=615,bbury=434,clip=]{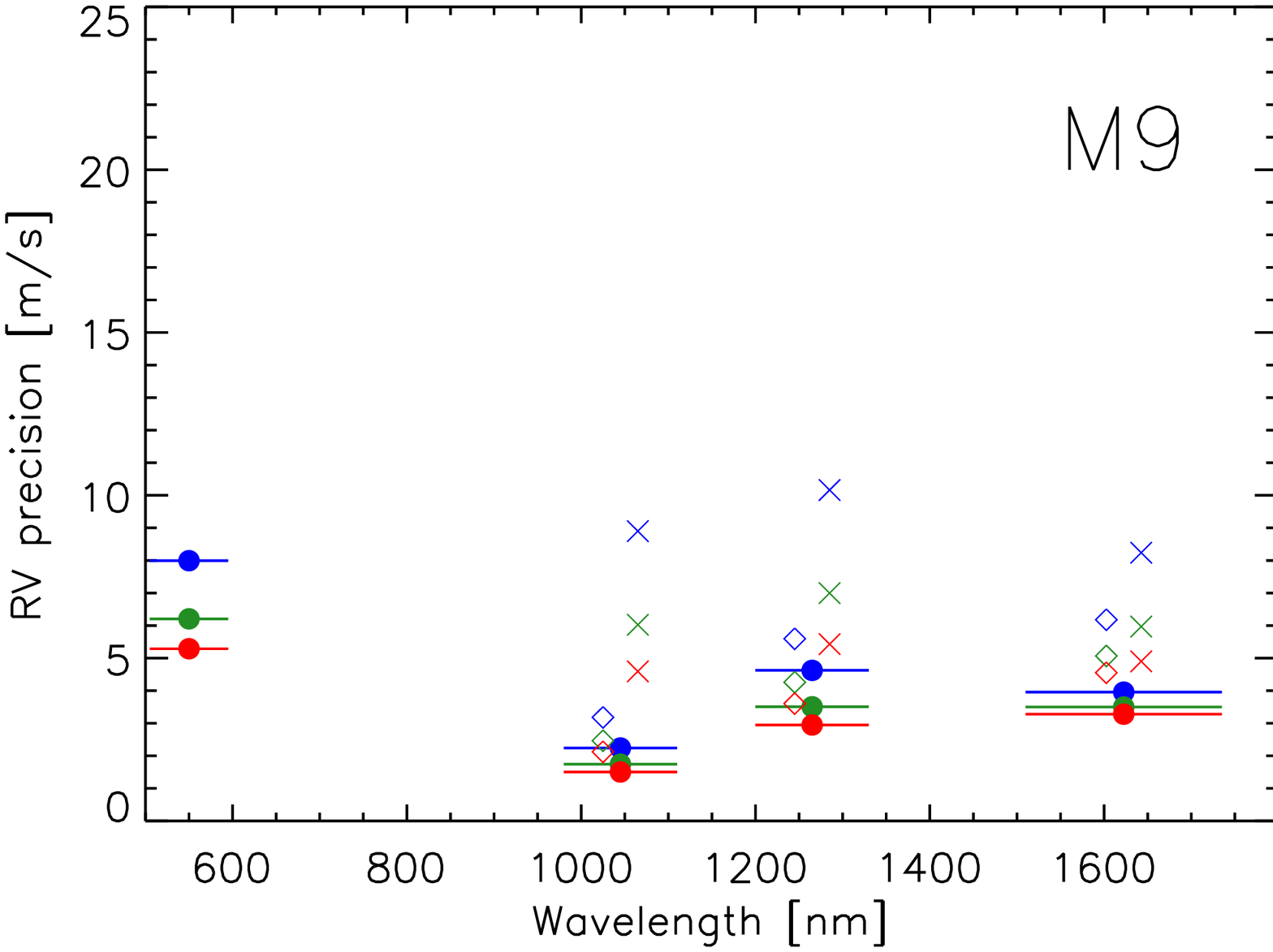}
  \caption{\label{fig:accuracies}Radial velocity precision for 3500\,K
    (M3, top panel), 2800\,K (M6, middle), and 2400\,K (M9, bottom).
    The situation is shown for three different spectral resolutions
    (red: $R = 100,000$; green: $R = 80,000$; blue: $R = 60,000$),
    $S/N$ is scaled according to the spectral resolution and assuming
    constant instrument efficiency (see Table\,\ref{tab:accuracies}).
    Horizontal lines show the spectral coverage used for the
    calculation. Filled circles show the best achievable precision
    assuming perfect wavelength calibration, i.e., the intrinsic
    stellar information content.  Open rhombs and crosses show the
    situation for wavelength calibration using ThAr lines and a
    hypothetical gas cell, respectively. In $V$, only the ideal case
    is shown because the wavelength calibration is not the limiting
    factor for the situations shown here.}
\end{figure}

The precision that can be achieved using (model) spectra of stars at
3500\,K (M3), 2800\,K (M6), and 2400\,K (M9) for different wavelength
regions are summarized in Table\,\ref{tab:accuracies} and shown in
Fig.\,\ref{fig:accuracies}. For each case, we first calculated the
intrinsic precision over the wavelength bin under consideration. As
explained in \S\ref{sect:precision}, we assume $S/N$ of 100 at
1\,$\mu$m at $R = 60,000$ and scale the signal quality according to
the spectral flux distribution and spectral resolution.  The
differences between radial velocity precisions at different wavelength
bands are dominated by the differences between the $S/N$ and between
the appearance of spectral features in these bands (see
Figs.\,\ref{fig:Mstars} and \ref{fig:YJH}). A secondary effect is the
length of the spectral range that differs between the bands, but it is
not always the band with the largest coverage that yields the highest
precision. We show the situation for three different values of $R =
100,000$, 80,000, and 60,000.

The $S/N$ given in Table\,\ref{tab:accuracies} varies according to the
number of photons per pixel, which decreases at higher spectral
resolution because the wavelength bins per pixel become smaller. The
$S/N$ is always comparable between the three nIR bands, but the
optical wavelength range provides $S/N$ that is about a factor of two
smaller in the M3, a factor of five smaller at M6, and a factor of ten
smaller in the M9 star.

In addition to the intrinsic precision, we show the precision
achievable if an imperfect wavelength calibration is considered. The
additional uncertainty due to ThAr or gas cell calibration (see
\S\ref{sect:calibration}) leads to somewhat higher limits that can be
achieved in a real observation. We show no ThAr or gas cell values for
the $V$-band because here the wavelength calibration is not the
critical factor for the situations investigated in this paper.

The question Fig.\,\ref{fig:accuracies} tries to answer is
what is the highest attainable precision of a radial
velocity measurement if a given star is observed at different
wavelength regions and spectral resolutions, under the assumption of
the same exposure time, telescope size, and instrument throughput for
all setups.

For an early-M star (M3), the highest precision is still reached in
the $V$-band, although the $Y$-band does not perform much worse. For
the given choice of parameters, the highest obtainable precision in
the $V$-band at $R = 100,000$ is roughly 2.5\,m\,s$^{-1}$, and in the
$Y$-band it is $\sim 3.8$\,m\,s$^{-1}$.  The $J$- and $H$-bands are
worse with only $\sim 16$\,m\,s$^{-1}$ and $8$\,m\,s$^{-1}$ precision,
respectively, at the highest resolution. In general, in the absence of
rotation, higher precision is obtained for higher spectral resolving
power\footnote{As an approximation, precision scales linearly with
  $S/N$ but quadratically with $R$. If a constant number of photons is
  assumed, $S/N$ scales down with $\sqrt{R}$, and as a result, the
  precision scales approximately with $R^{3/2}$.}. We discuss the
limits to the precision in rotating stars in $\S$\ref{sect:Rotation}.
A remarkable feature of our precision calculations for $T = 3500$\,K
is that although the flux level in the visual wavelength range is much
lower than the flux around 1\,$\mu$m and redder, the radial velocity
precision is not worse.  This is because the optical spectrum of an
early-M dwarf is extremely rich in narrow features. At nIR
wavelengths, the number of features is much lower so that the
attainable precision is lower, too. The same explanation holds for the
comparison between the nIR $Y$-, $J$-, and $H$-bands. The low
precision obtainable in the $J$-band is due to the lack of sharp and
deep spectral features in that wavelength range (compare
Figs.\,\ref{fig:Mstars} and \ref{fig:YJH}).

At lower temperature, $T = 2800$\,K (M6), the overall precision (at
the same $S/N$, that means in general after longer exposures) has
gained in the nIR-bands in comparison to the optical, because now the
$S/N$ is much higher at nIR regions.  The $Y$- and $J$-band precisions
improve a lot in comparison to the M3 star, which is also due to the
appearance of FeH bands \citep[see][]{Cushing05}. The $H$-band
precision at 2800\,K is comparable to the 3500\,K spectrum.  Now, the
$V$-band performs worse than the $Y$-band, but still it yields a
remarkably high precision: Although the flux level in the $V$-band is
about an order of magnitude below the flux at nIR wavelengts, the
richness in sharp features can almost compensate for this. The
$V$-band precision is only about 30\,\% worse than the $Y$-band
precision, and it still yields much better precision than the $J$- and
the $H$-bands.

Finally, in the M9 star at $T = 2400$\,K, all three nIR bands
outperform the $V$-band because of the high flux ratio between the nIR
and the optical range. Still, the $Y$-band provides the highest
precision about a factor of two better than the $J$- and the $H$-band.

We consider now the effects of limited precision in the wavelength
calibration using the ThAr lamp or a gas cell (shown in in
Fig.\,\ref{fig:accuracies} as open rhombs and crosses, respectively).
The ThAr calibration apparently can provide a very reliable
calibration that introduces only a few percent loss in precision. Of
course, in a real spectrograph, effects like wavelength stability over
time additionally limit the precision that can be reached
\citep{Lovis07}. This effect has not been taken into account.
Nevertheless, our calculations show that enough suitable ThAr lines
are available and that a wavelength solution at nIR wavelengths ($Y$
-- $H$-bands) based on ThAr lines is a reliable calibration that can
in principle be expected to work almost as successfully as in the
optical wavelength range.  In contrast to that, the calibration using
the virtual gas cell as a reference yields much worse a result, in
particular at short wavelengths like in the $Y$-band. In order to make
the gas-cell calibration provide the same accuracy as a calibration
using a ThAr lamp (in a stabilized spectrograph), a gas is needed that
provides more and deeper lines than NH$_3$ provides in the $K$-band.
So far, all gases known provide many fewer lines so that currently
achievable precision turns out to be significantly below than what can
be achieved with ThAr. We note that in order to make the gas cell
calibration work, the spectrum must have a minimum $S/N$ allowing a
reliable reconstruction of the instrument profile. This means a
typical minimum $S/N$ of $\sim 100$ is required for the gas cell
method. Thus, using a gas cell for the wavelength calibration,
low-$S/N$ spectra as in the M6 and M9 cases in the $V$-band considered
above could not be used.

\subsection{The influence of rotation}
\label{sect:Rotation}

\subsubsection{Distribution of rotation in M dwarfs}

\begin{figure}
  \plotone{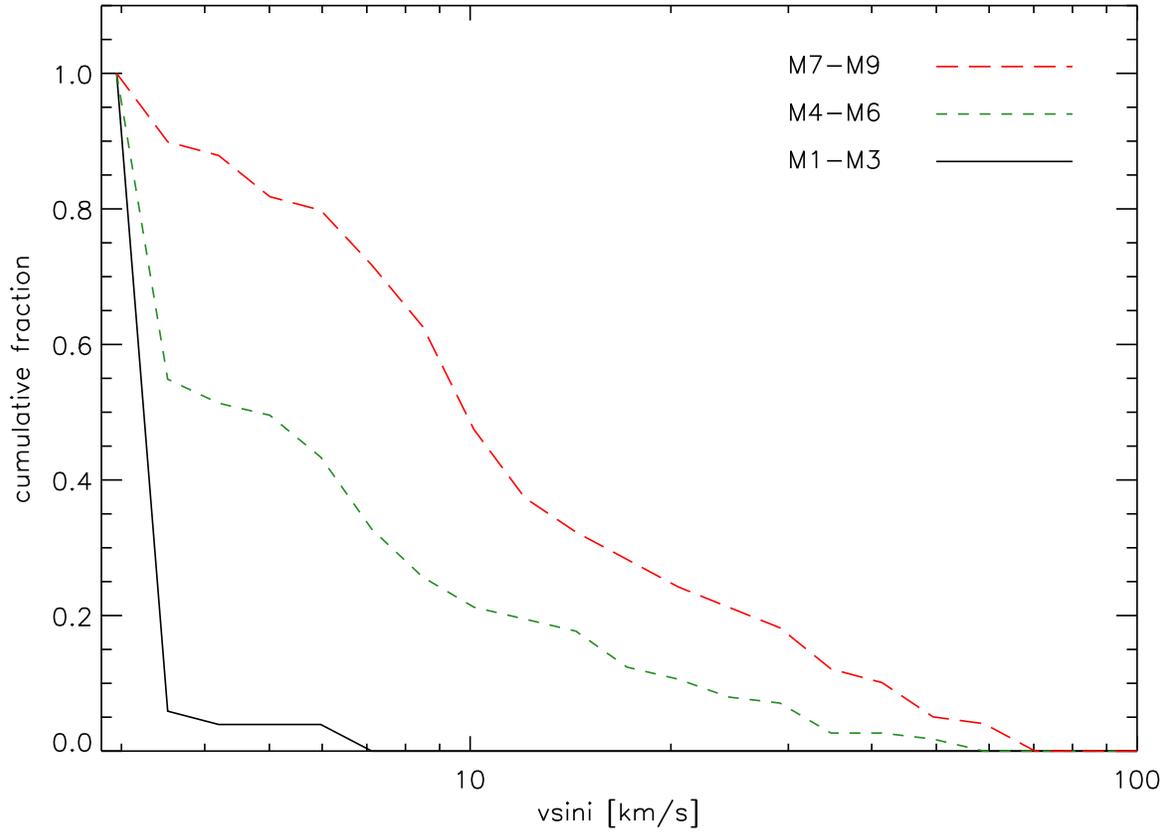}
  \caption{\label{fig:rotfrac}Distribution of rotational velocities
    among field M dwarfs. Cumulative plot showing the fraction of
    early-M (M1--3), mid-M (M4--M6), and late-M (M7--M9) stars
    rotating faster than a given value of $v\,\sin{i}$. Data are from
    \cite{Delfosse98, Mohanty03, Reiners08} and \citet{Reiners09}. A
    lower detection limit of $v\,\sin{i} = 3$\,km\,s$^{-1}$ is assumed
    for all data.}
\end{figure}

Higher resolution spectra only offer an advantage for radial velocity
measurements if the stars exhibit sharp lines \citep[see
also][]{Bouchy01}. Field mid- and late-M dwarfs can be very rapid
rotators with broad and shallow spectral lines. For example,
\citet{Reiners08} show that rapid rotation is more frequent in cooler
M dwarfs. We have collected measurements on rotational velocities from
\citet{Delfosse98, Mohanty03, Reiners08} and \citet{Reiners09}, and we
show the cumulative distributions of $v\,\sin{i}$ for early-, mid-,
and late-M stars in Fig.~\ref{fig:rotfrac}. All early-M dwarfs (M1 --
M3) in the samples are rotating slower than $v\,\sin{i} =
7$\,km\,s$^{-1}$, which means that early-M dwarfs are very good
targets for high precision radial velocity surveys. Among the mid-M
dwarfs (M4 -- M6), about 20\,\% of the stars are rotating faster than
$v\,\sin{i} = 10$\,km\,s$^{-1}$, and this fraction goes up to about
50\,\% in the late-M dwarfs (M7 -- M9).

\subsubsection{ Rotational broadening and radial velocity precision}

\begin{figure}
  \plotone{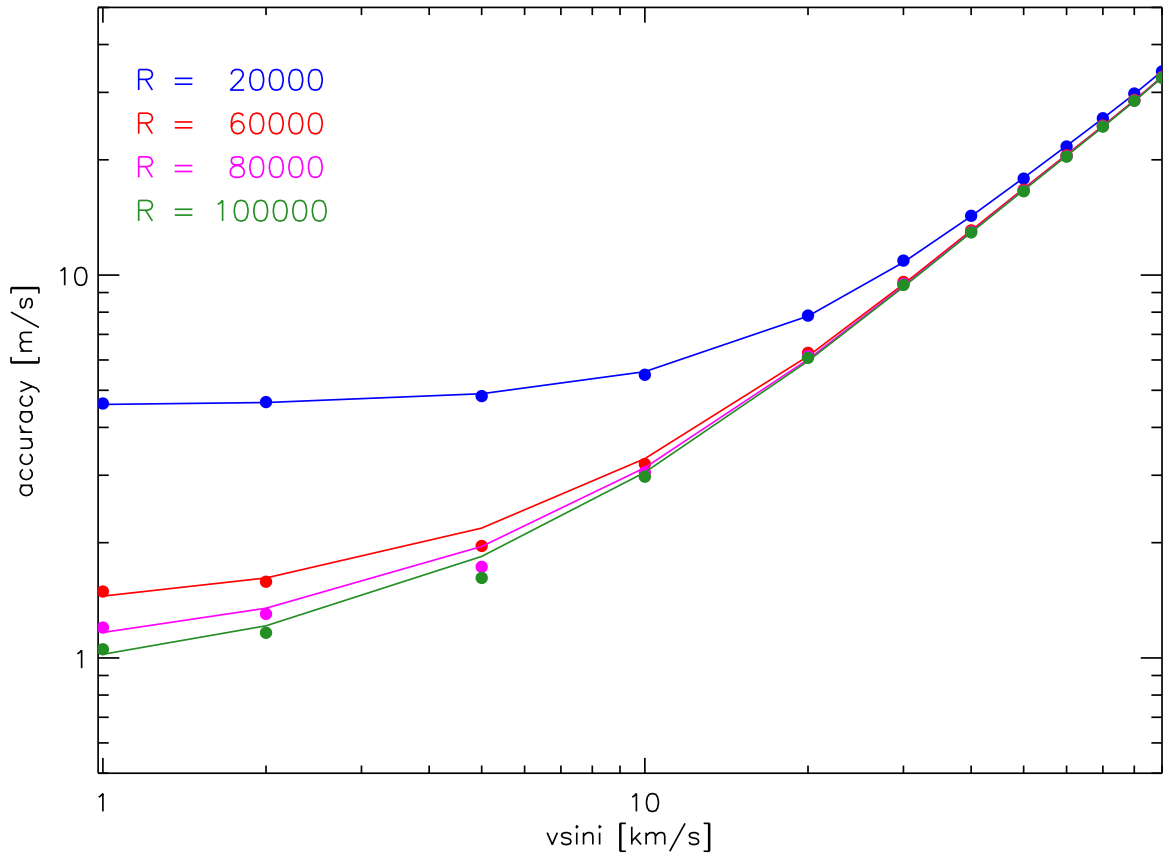}
  \caption{\label{fig:accvsini}Radial velocity precision as a function
    of rotational velocity for four different resolving powers. The
    assumed model is a 3000\,K star and the precision was calculated
    in the $Y$-band.}
\end{figure}

We show in Fig.~\ref{fig:accvsini} the $Y$-band precision of radial
velocity measurements that can be achieved in a 3000\,K star (M5) as a
function of projected rotational velocity at different spectral
resolutions ($R$ = 20,000, 60,000, 80,000, and 100,000). All
assumptions about $S/N$ and $R$ are as explained above.

As expected, at high rotation velocities ($v\,\sin{i} >
30$\,km\,$^{-1}$), the precision achieved with spectrographs operating
at different $R$ are hardly distinguishable. Only if the line width of
the rotating star is narrower than the instrumental profile does
higher resolution yield higher precision. However, for $R > 60,000$,
the difference in precision is relatively small even at very slow
rotation. At a rotation velocity of $v\,\sin{i} = 10$\,km\,s$^{-1}$,
the precision is roughly a factor of 3 lower than the precision in a
star with $v\,\sin{i} = 1$\,km\,s$^{-1}$, and $v\,\sin{i} =
6$\,km\,s$^{-1}$ brings down the precision by about a factor of 2
compared to $v\,\sin{i} = 1$\,km\,s$^{-1}$.

\section{Radial velocity variations from starspots}

In this Section, we investigate the influence of starspots on the
measurement of radial velocities. A similar study was performed by
\citet{Desort07}, but for sun-like stars in the optical wavelength
regime. We extend this study to cooler stars and nIR wavelengths.

\subsection{Spot properties}

Radial velocity variations can be mimicked by surface features
corotating with the star. We know from the Sun that magnetic activity
can cause spots that are significantly cooler than the photosphere,
and much larger spots are found on stars more active than the Sun (in
general, rapidly rotating stars are more active). However, the
temperature differences between starspots and the corresponding
``quiet'' photosphere remains a rather unknown parameter, especially
for very active stars.

The coolest temperatures in large sunspots can be $\sim\,2000$\,K
lower than the temperature in the surrounding photosphere
\citep{Solanki03}.  Observed as a star, however, such differences
would still be difficult to observe because sunspots cover only a very
small fraction of the surface ($< 1\%$). \cite{ONeal01, ONeal04}
reported spot temperatures up to 1500\,K below photospheric
temperatures in active G- and K-type stars. These spots cover between
10 and 50\,\% of the stellar surface.  \citet{Strassmeier98} reported
spot temperatures in an active K dwarf that are $\sim 800$\,K below
the effective photospheric temperature based on Doppler imaging.

From the available observations, no general prediction on the
occurrence and properties of spots on stellar surfaces is possible. In
particular, no observational information is available on spot
distributions on M dwarfs, which have atmospheres significantly
different than those of sun-like stars, and may also have different
magnetic field topologies \citep{Donati08, RB09}. Therefore, we
investigate a set of different photosphere and spot temperature
combinations in the following.

\begin{figure}
  \plotone{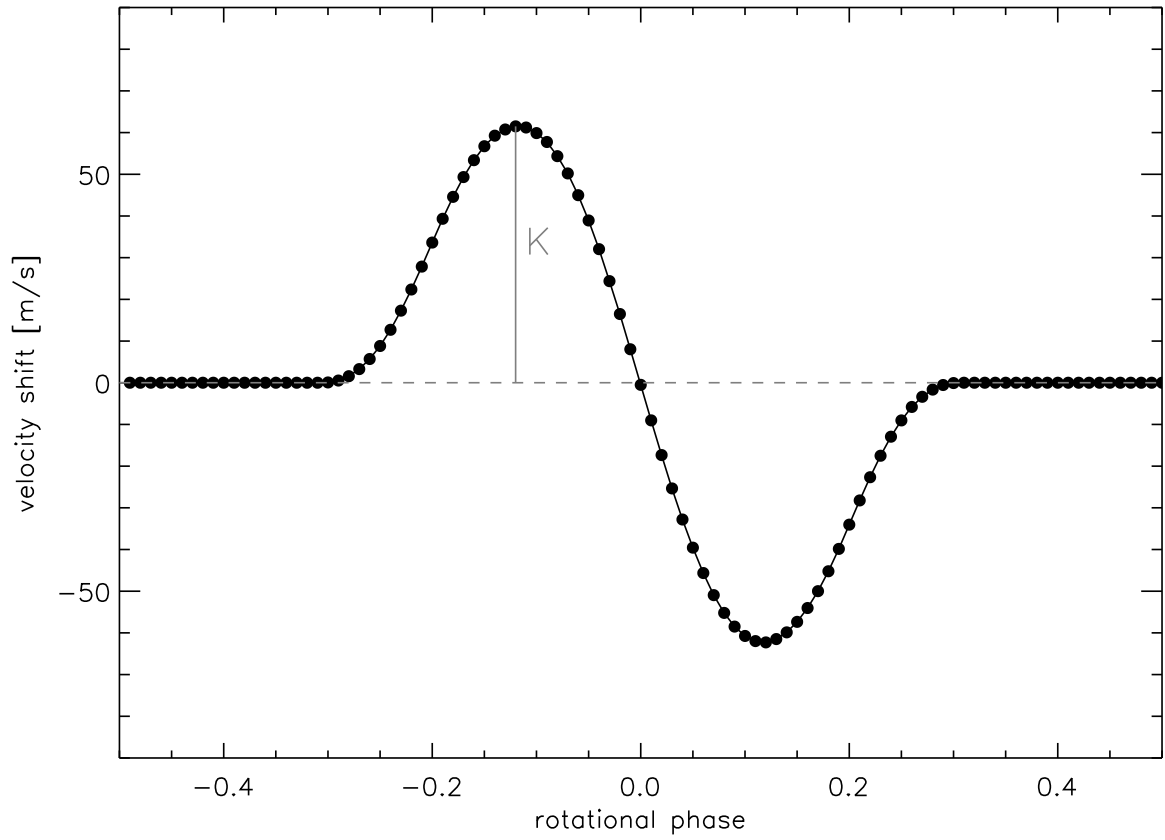}
  \caption{\label{fig:spotexample}Example of the apparent radial
    velocity shift induced by a single cool spot as a function of
    rotational phase. Our assumed definition of the radial velocity
    amplitude, $K$, is indicated.}
\end{figure}

Before we investigate apparent radial velocity shifts using a detailed
model (\S4.2), we consider a ``toy model'' of an ideal spectral line
composed of signal from two regions: the ``quiet'' surface of a
rotating star, and a co-rotating, cool spot. To demonstrate the
influence of the temperature contrast and differing spectral features
for a spot, we generate the line profiles sampled over a complete
rotational period and compute the apparent radial velocity shift for
each rotational phase by fitting them with a Gaussian. An example
radial velocity curve is given in Fig.~\ref{fig:spotexample}. We
report on the amplitude $K$ of the radial velocity curve occurring
during a full rotation\footnote{\citet{Desort07} report peak-to-peak
  amplitude, which is twice the value we are using.}.

\subsubsection{Contrast}
\label{sect:Contrast}

\begin{figure}
  \plotone{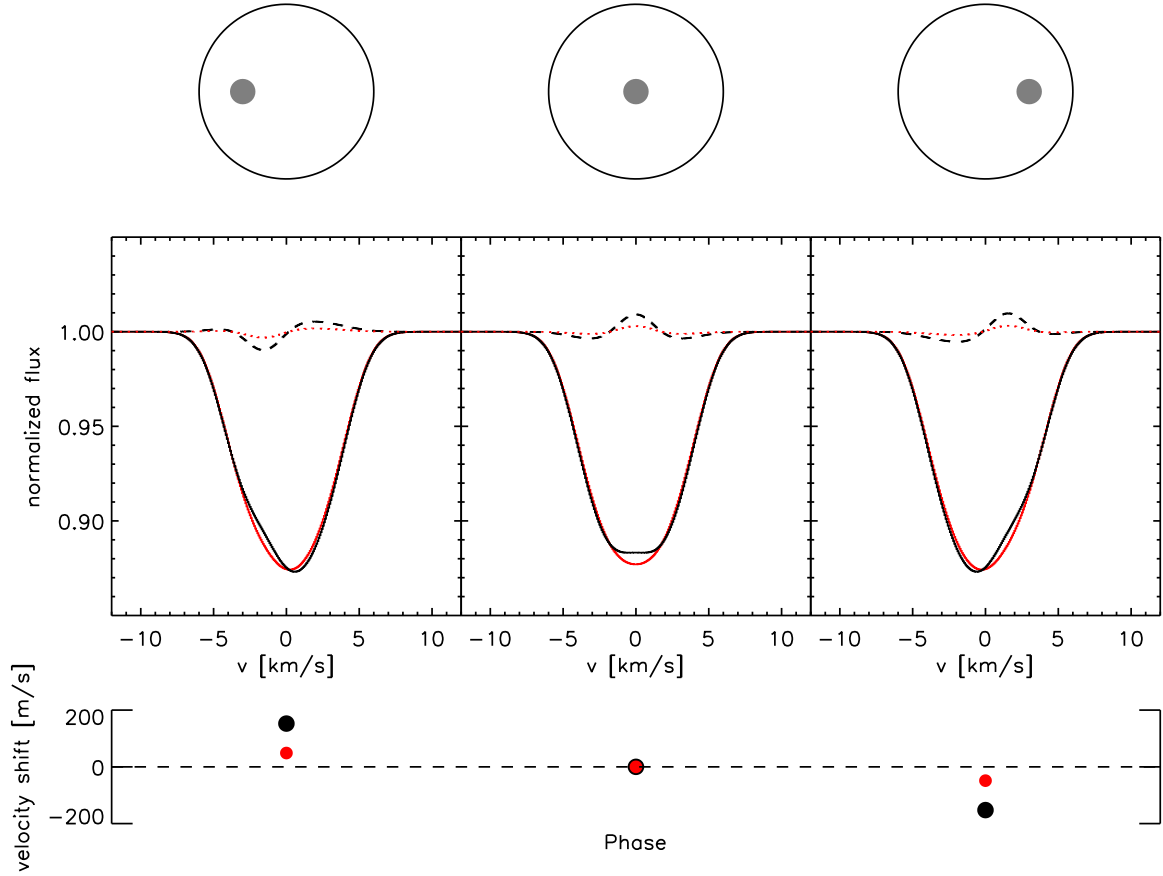}
  \caption{\label{fig:spotillu}Illustration of the apparent radial
    velocity shift due to a single spot for two contrast values. From
    left to right, three different phases are shown, the top panel
    illustrates the location of the spot at each phase. The center
    panel shows line profiles (solid lines) and residuals (dashed
    lines; residual between the profile of the quiet photosphere and
    the profile of the spotted star) for a star with a photospheric
    temperature of $T = 3700$\,K rotating at $v\sin{i} =
    5$\,km\,s$^{-1}$. Black and red lines indicate different spot
    temperatures (black: $T_{\rm spot} = 0$\,K, red: $T_{\rm spot} =
    3500$\,K,). The lower panel shows the measured radial velocity
    shift.}
\end{figure}

One crucial parameter for the influence of a starspot on a spectral
line, and thus the star's measured radial velocity, is the flux
contrast between the quiet surface and the spot at the wavelength of
interest. In general, one expects the radial velocity amplitude due to
a spot to be smaller with lower contrast and vice versa. We illustrate
the effect of a cool spot on the line profile and the measured radial
velocity shift in Fig.~\ref{fig:spotillu}, where the situation for a
spot covering 2\,\% of the visible surface is shown for two different
temperatures (contrasts).

To investigate the influence of contrast over a wide range of
parameters we used the toy model described above.  We assumed a
blackbody surface brightness according to a surface temperature $T_0$,
and we subtracted the amount of flux that originates in a spot
covering 2\,\% of the stellar surface. We then added the flux
originating in the spot at a temperature $T_1$. This spot has the same
line profile as the photosphere, which means that we assume a constant
line profile for the whole star. For our example, we chose a
rotational velocity of $v\,\sin{i} = 2$\,km\,s$^{-1}$.

\begin{figure}
  \plotone{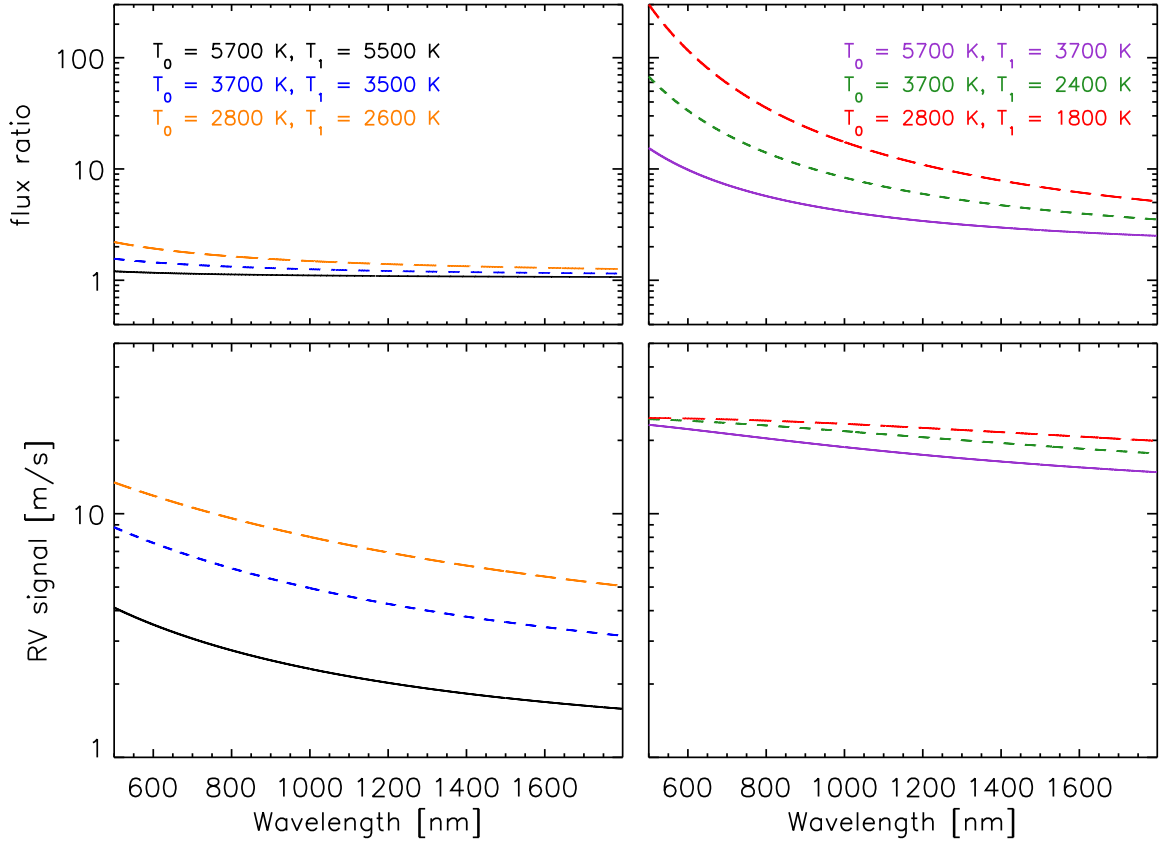}
  \caption{\label{fig:contrast}Apparent radial velocity shift by a
    single spot for different temperature contrasts calculated with
    our ``toy model''. Upper panel: Flux ratio due to a cool spot with
    temperature $T_1$ on a surface at temperature $T_0$, the contrast
    follows the ratio of the black-body distributions of the flux.
    Lower panel: Radial velocity signal induced by the spot during a
    rotation of the star at $v\sin{i} = 2$\,km\,s$^{-1}$. The left
    panel shows the situation of small contrast ($T_0 - T_1 =
    200$\,K), in the right panel the contrast is larger ($(T_0 -
    T_1)/T_0 = 0.35$). }
\end{figure}

In Fig.~\ref{fig:contrast}, we show the wavelength-dependent flux
ratio between the quiet photosphere and the spot (the contrast, upper
panel), and the resulting apparent radial velocity shift from the toy
model calculations (lower panel). We show cases for three photosphere
temperatures $T_0$ (5700\,K, 3700\,K, and 2800\,K), for each $T_0$ we
show cases with a small temperature difference, $T_0 - T_1 = 200$\,K,
and with a larger temperature difference of $(T_0 - T_1)/T_0 = 0.35$.

For small temperature differences (left panel of
Fig.~\ref{fig:contrast}), the flux ratio between photosphere and spot
decreases by approximately a factor of two in the range 500 -- 1800\,nm,
while the radial velocity (RV) signal decreases by roughly a factor
between two and three. The RV signal is higher for the lowest $T_0$
because the relative difference between $T_1$ and $T_0$ is much larger
than in the case of $T_0 = 5700$\,K.

The cases of large temperature contrast (right panel in
Fig.\,\ref{fig:contrast}) produce flux ratios $>100$ in the coolest
star at 500\,nm.  At 1800\,nm the flux ratio decreases to a value
around 5, i.e., a factor of 20 lower than at 500\,nm. This is much
larger than for the cases with small temperature contrast, where the
flux ratio only decreases by a factor of two or less. On the other
hand, the RV signal does not change as dramatically as the flux ratio.
For large temperature contrasts, the absolute values of the RV signal
are larger than in the case of low contrast, but the slope of RV with
wavelength is much shallower; it is well below a factor of two in all
three modeled stars.  The explanation for this is that the large
contrast in flux ratio implies that the spot does not contribute a
significant amount to the full spectrum of the star at any of these
wavelengths.  Therefore, a relatively small change of the flux ratio
with wavelength has no substantial effect on the RV signal. If, on the
other hand, the flux ratio is on the order of two or less, a small
decrease in the ratio with wavelength means that the larger
contribution from the spot can substantially change the RV signal.
Thus, a significant wavelength dependence for an RV signal induced by
a cool spot can only be expected if the temperature difference between
the quiet photosphere and the spot is not too large.

\subsubsection{Line profile in the spot}

Line profile deviations that cause radial velocity variations do not
solely depend on temperature contrast but also on the dependence of
spectral features on temperature; effective temperature generally
corresponds to a different spectrum and does not just introduce a
scaling factor in the emitted flux. For example, a spectral line
variation, and hence a radial velocity shift, can also appear at zero
temperature contrast if the spectral line depths differ between the
spot and the photosphere \citep[as for example in hot stars with
abundance spots;][]{Piskunov93}.

\begin{figure}
  \plotone{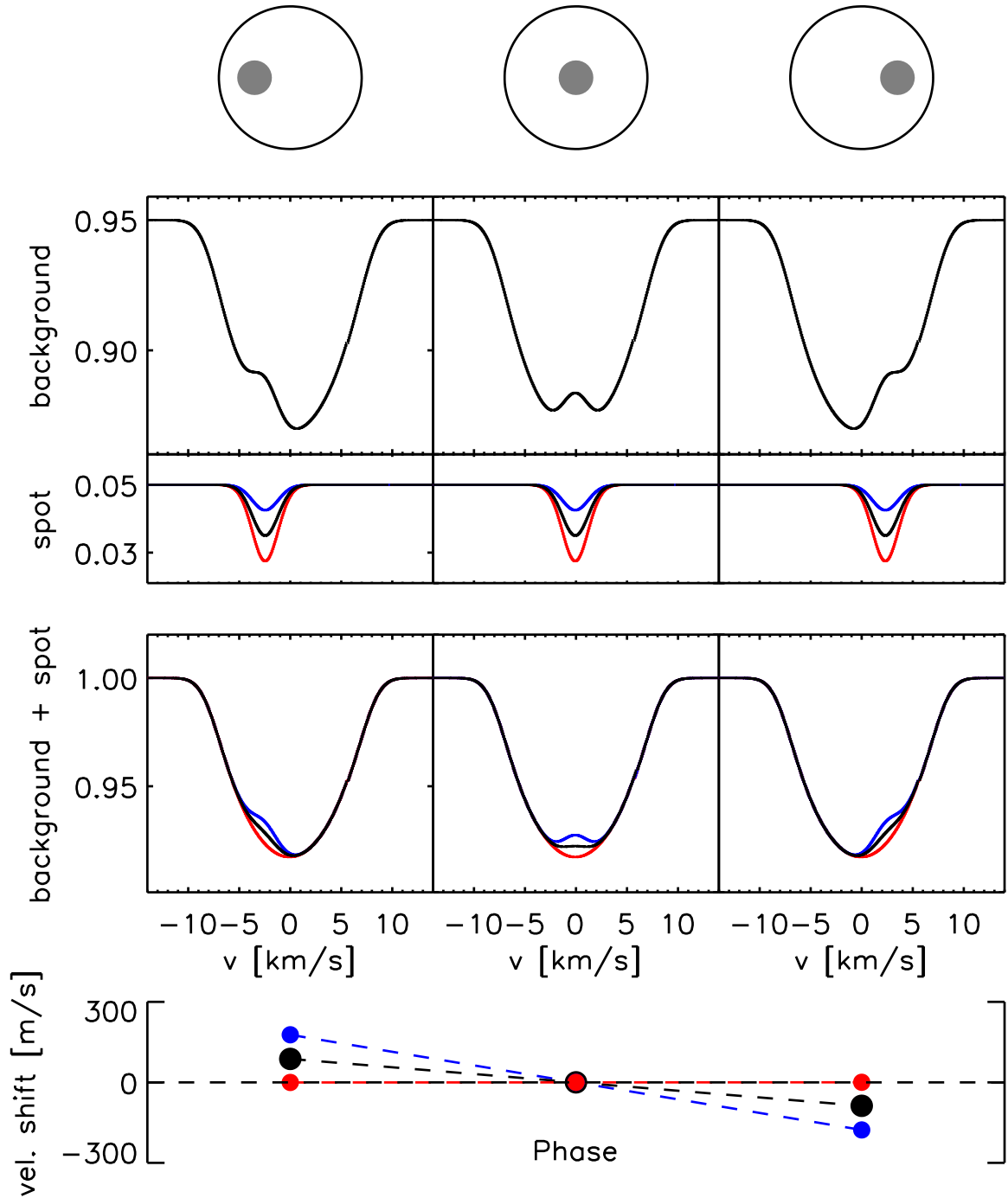}
  \caption{\label{fig:spotdepths}Illustration of apparent radial
    velocity shift by a single spot for different spot line
    intensities (toy model). The upper panel of the line profile plots
    gives the line profile contribution from the background
    photosphere. The middle panel gives the three different line
    profile contributions from a cool spot. The lower panel shows the
    sum of the two line profile contributions, i.e., the final
    spectrum of the spotted star for three cases of different spot
    line depths. The flux scale is normalized so that the total
    spectrum has a continuum value of 1, note the background continuum
    is at a value of 0.95 and the spot continuum at 0.05. The black
    line shows the case where the spot and background photosphere
    lines are identical. The bottom plot shows the resulting radial
    velocities for the different line profile combinations.}
\end{figure}

In Fig.~\ref{fig:spotdepths}, we consider a similar situation as in
Fig.~\ref{fig:spotillu}. Here, the temperature contrast between spot
and photosphere is held constant but we show three cases in which the
depths of the spectral line originating in the spot are different. The
three spot profiles are chosen so that the line depths are 0.5, 1.0,
and 1.5 times the line depth of the photospheric line.
Fig.~\ref{fig:spotdepths} illustrates that if spectral features are
weaker in the spot than in the surrounding photosphere, the radial
velocity shift (bottom panel) is larger than in the case of identical
line strengths. If, on the other hand, the spectral features become
stronger, as for example in some molecular absorption bands, the spot
signature weakens and the radial velocity distortion is smaller. In
our example of a stronger spot feature, this effect entirely cancels
out the effect of the temperature contrast, so that the radial
velocity signal of the spot is zero although a cool spot is present.

\subsection{Spot simulations}

After considering the general effects of starspots in the last
Section, we now discuss results of a more sophisticated simulation.
Here, we calculate a full spectrum by integrating over the surface of
a spotted star using spectra from a model-atmosphere code. The
resulting radial velocity shift is estimated by cross-correlation
against a spectrum unaffected by spots.

\subsubsection{Line profile integration}

Our model spectra for a spotted star were calculated using a discrete
surface with \mbox{$500 \times 500$~elements} in longitude and
latitude arranged to cover the same fraction of the surface each.  All
surface elements are characterized by a `local' temperature; unspotted
areas are assigned to the photospheric temperature, and spotted areas
contribute spectra pertaining to atmospheres with lower temperatures.
The associated spectra $f(\lambda, T)$ were generated with PHOENIX for
all temperatures used. Depending on the rotational phase $p$, the
visibility $A_i$ of each surface element $i$ is calculated,
considering projection effects. We determine the radial velocity shift
$v_{{\rm rad},i}$ for each surface element due to the stellar
rotation.  The resulting model spectrum $f_p(\lambda)$ for the spotted
star is
\begin{equation}
  f_p(\lambda) = \frac{\sum \limits_{i=1}^N f(\lambda, T_i, v_{{\rm rad},i}) A_i}{\sum \limits_{i=1}^N A_i},
\end{equation}
where $N$ is the total number of elements.

In all of our calculations the model star has an inclination of
\mbox{$i \, = \, 90\degr$}, and for simplicity we assume a linear limb
darkening coefficient \mbox{$\epsilon \, = \, 0$} (no limb darkening).
Using no limb darkening slightly overestimates the radial velocity
signal but captures the qualitative behaviour that we are interested
in.  Stellar spots are considered to be circular and located at
$0\degr$~longitude and $+30\degr$~latitude.

We calculated the RV signal introduced by a temperature-spot on a
rotating star. We chose the same star/spot temperature pairs as in the
contrast calculations in the foregoing Section, but we used detailed
atmosphere models and PHOENIX spectra to calculate the RV signal over
the wavelength area 500 -- 1800\,nm. That means, our calculations
include the effects of both the contrast \emph{and} differences in the
spectral features between atmospheres of different temperatures.

\subsubsection{Results from spot simulations}

\begin{figure}
  \plotone{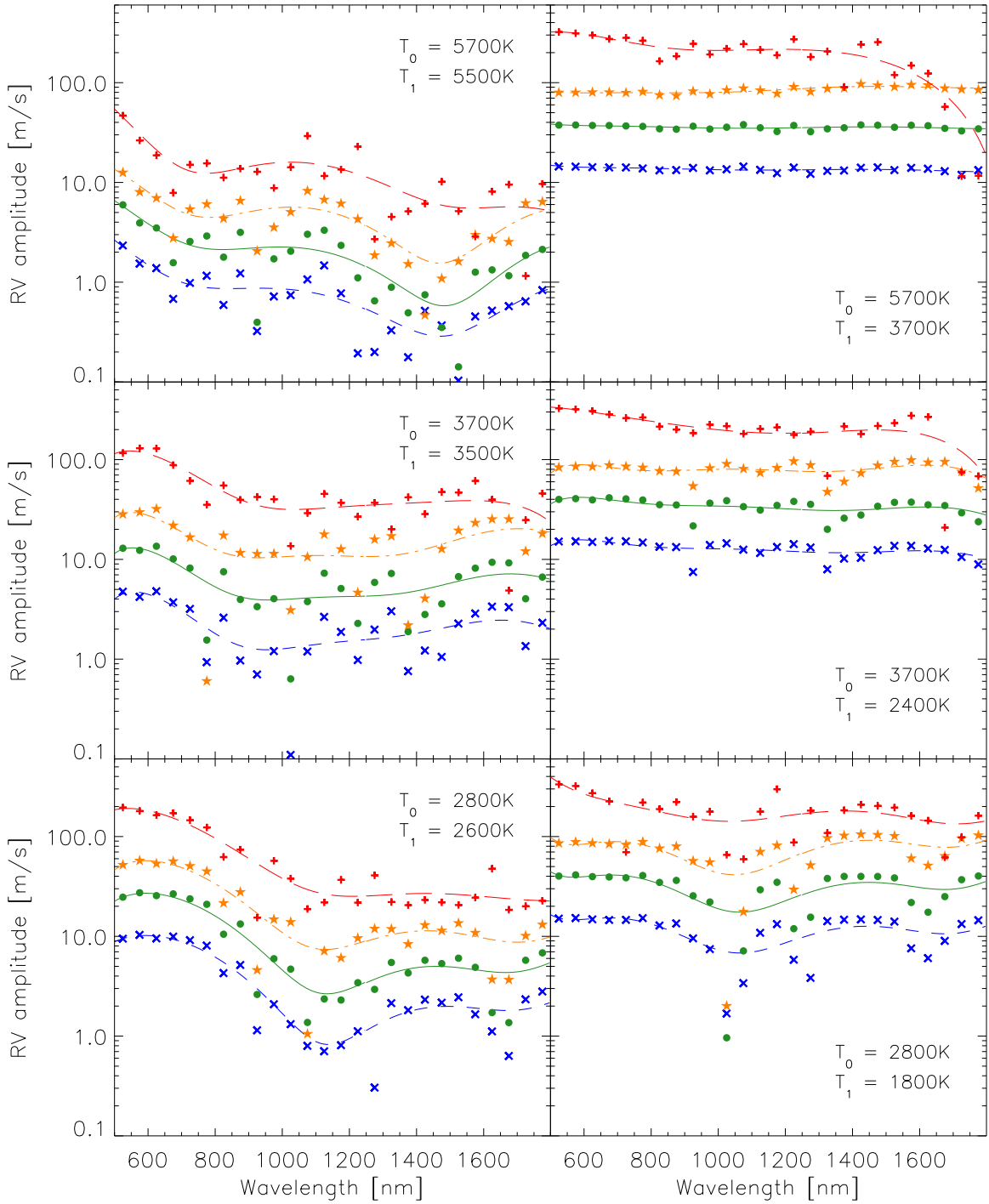}
  \caption{\label{fig:spotsimulation}Simulations of RV amplitude as a
    function of wavelength for different temperature combinations and
    different rotation velocities. Individual points indicate the RV
    amplitude at one wavelength chunk, blue crosses are for
    $v\,\sin{i} = 2\,$\,km\,s$^{-1}$, green circles: $v\,\sin{i} =
    5$\,\,km\,s$^{-1}$; orange stars: $v\,\sin{i} = 10$\,km\,s$^{-1}$;
    red plusses: $v\,\sin{i} = 30$\,km\,s$^{-1}$. Temperatures of the
    photosphere ($T_0$) and the spot ($T_1$) are shown in the panels.
    The spot has a size of 1\,\% of the projected surface. Individual
    points are connected by a polynomial fit to guide the eye. }
\end{figure}

The results of our calculations are shown in
Fig.\,\ref{fig:spotsimulation}. As in Fig.~\ref{fig:contrast}, we show
six different stars, the temperature combinations are identical.  The
spot size is 1\,\% of the projected surface. We compute RV amplitudes
for 50\,nm wide sections. For each model star, we show four cases for
rotational velocities of $v\,\sin{i} = 2, 5, 10$, and
30\,km\,s$^{-1}$.

In general, the trends seen in the detailed model are consistent with
our results from the toy model of a single line taking into account
only the flux ratio between photosphere and spot. As long as the flux
ratio is relatively small (200\,K, left panel in
Fig.~\ref{fig:spotsimulation}), the RV amplitude strongly depends on
wavelength due to the wavelength dependency of the contrast. The
strongest gradient in RV amplitude occurs between 500 and 800\,nm; the
RV signal decreases by almost a factor of 10 in the lowest temperature
model where the flux gradient is particularly steep over this
wavelength range. The decrease occurs at somewhat longer wavelengths
in the cooler model stars than in the hotter models. In the model
stars with a high flux ratio between photosphere and spot (right panel
in Fig.\,\ref{fig:spotsimulation}), the variation in RV amplitude with
wavelength is very small. The very coolest model shows a few regions
of very low RV amplitude, but the general decline is not substantial.

The absolute RV signal in a Sun-like star with $T_0 = 5700$\,K at
500\,nm comes out to be $\sim 40$\,m\,s$^{-1}$ at a spot temperature
of $T_1 = 3700$\,K on a star rotating at $v\,\sin{i} =
5$\,km\,s$^{-1}$. This is consistent with the result of
\citet{Desort07}, who reported a peak-to-peak amplitude of $\sim
100$\,m\,s$^{-1}$, i.e., an ``amplitude'' of 50\,m\,s$^{-1}$, in a
similar star. Over the wavelength range they investigated (roughly 500
-- 600\,nm), Desort et al. found that the RV amplitude decreases by
about 10\,\%. We have not calculated the RV amplitude over bins as
small as the bins in their work and our calculation only has two bins
in this narrow wavelength range. However, we find that the decrease
\citet{Desort07} reported between 500\,nm and 600\,nm, is not
continued towards longer wavelengths. In our calculations, the RV
amplitude does not decrease by more than $\sim 20\,\%$ between 500 and
1800\,nm.  Similar results apply for higher rotation velocities and
lower photosphere temperatures.

The models with lower flux ratio, $T_0 - T_1 = 200$\,K, show more of
an effect with wavelength, although the absolute RV signal is of
course smaller. The RV amplitude in a star with $T_0 = 3700$\,K, a
spot temperature of $T_1 = 3500\,$K, and $v\,\sin{i} =
5$\,km\,s$^{-1}$ is slightly above 10\,m\,s$^{-1}$ at 500\,nm and
$\sim 4$\,m\,s$^{-1}$ at 1000\,nm. Above 1000\,nm, no further decrease
in RV amplitude is observed. The behavior, again, is similar in other
stars with low flux ratios and with different rotation velocities.

For the cool models with large temperature differences, the individual
RV amplitudes show relatively large scatter between individual
wavelength bins. We attribute this to the presence of absorption
features in some areas, while other areas are relatively free of
absorption features. The temperature dependence of the depth of an
absorption feature is important for the behavior of the spot signature
in the spectral line. An example of this effect can be observed around
1100\,nm, where the spectrum is dominated by absorption of molecular
FeH that becomes stronger with lower temperature.

\subsubsection{Comparison to LP 944-20}

We can compare our simulations to the optical and nIR radial velocity
measurements in LP~944-20 reported by \citet{Martin06}. At optical
wavelengths, they found a periodical radial velocity variation of $K =
3.5$\,km\,s$^{-1}$, while in the nIR they could not find any
periodical variation and report an rms of 0.36\,km\,s$^{-1}$. The
approximate effective temperature of an M9 dwarf like LP~944-20 is
around 2400\,K, we compare it to our coolest model that has a
temperature of $T_0 = 2800$\,K. The radial velocity amplitude of
3.5\,km\,s$^{-1}$ at visual wavelengths is much higher than the
results of our simulations, but this can probably be accounted for by
the different size of the surface inhomogeneities (only 1\% in the
simulations)\footnote{Using our ``toy model'' we estimate that a spot
  covering $\sim10$\,\% of the surface can generate a velocity
  amplitude of a few km\,s$^{-1}$.}.

The observations of \citet{Martin06} suggest a ratio between optical
and nIR radial velocity variations larger than a factor of 10. The
largest ratio in our set of simulations in fact is on that order; the
radial velocity jitter in our model with $T_0 = 2800$\,K and $T_1 =
2600$\,K diminishes by about a factor of ten between 600\,nm and
1200\,nm.  Extrapolating from the results of the hotter models, in
which the ratio between visual and nIR jitter is smaller, we cannot
exclude that this ratio may become even larger in cooler stars.  Our
model with larger temperature contrast ($T_0 = 2800$\,K, $T_1 =
1800$\,K) produces a smaller ratio between visual and nIR jitter.
Thus, our simulations are not in contradiction to the results reported
by \citet{Martin06}. A ratio of ten or more in jitter between optical
and nIR measurements seems possible if the temperature contrast is
fairly small (100--200\,K).

We note that no radial velocity variations were actually detected in
LP~944-20 in the nIR, which means that at the time of the nIR
observations, it simply could have experienced a phase of lower
activity and reduced spot coverage.  In order to confirm the effects
of a wavelength-dependent contrast on radial velocity measurements,
observations carried out simultaneously at visual and nIR wavelenghts
are required.

\citet{Martin06} propose that weather effects like variable cloud
coverage may be the source for the radial velocity jitter in the
visual and for the wavelength-dependent contrast. This would probably
mean very small temperature contrast but a strong effect in wavelength
regions with spectral features that are particularly sensitive to
dust. Our simulations do not predict the wavelength dependence of pure
dust clouds, but at this point we see no particular reason why the
jitter from purely dust-related clouds should be much stronger in the
visual than in the nIR. To model this, a separate simulation would be
needed taking into account the effects of dust on the spectra of
ultracool dwarfs.

\section{Summary and Conclusions}

We have investigated the possibility of measuring radial velocity
variations of low-mass stars at nIR wavelengths ($Y, J, H$). The
spectral flux distribution of red stars favors long wavelengths
because higher $S/N$ can be achieved in comparison to optical
wavelengths. On the other hand, the spectral information content of
the spectra (presence of sharp and strong spectral features) is lower
at longer wavelengths, and the efficiency of calibration methods is
not well established.

For early M dwarfs, nIR radial velocities do not offer any advantage
in terms of photon-limited precision. Indeed, the fact that
measurement methods in the optical are much more advanced than those
in the nIR means that there is not really any motivation of nIR radial
velocities from this perspective. On the other hand, at mid-M spectral
type, the achievable precision becomes higher in the nIR around
spectral type M4--5; $Y$-band observations can be expected to achieve
a radial velocity precision higher than observations at optical
wavelengths. At late-M dwarfs, the $Y$-band outperforms the $V$-band
precision by about a factor of 4--5. Observations in the $J$- and
$H$-bands are a factor of 2--5 worse than the $Y$-band across the M
star spectral types. They are only superior to the $V$-band
observations in very-late-M dwarfs.

Our investigation into the effects of activity on radial velocity
measurements showed that a crucial parameter for the wavelength
dependence of jitter is the temperature contrast between the spot and
the photosphere. If the spot temperature is only a few hundred Kelvin
below the photospheric temperature, the induced radial velocity signal
is on the order of several m\,s$^{-1}$ in the optical and becomes
weaker towards longer wavelengths. Note that the absolute size of this
effect depends on the size of the spot (1\,\% in our simulation) and
will grow with larger spot size. High temperature contrast, on the
other hand, causes a much larger radial velocity signal that only
weakly depends on wavelength. For example, in M stars with spots only
$\sim 200$\,K cooler than the photosphere, the jitter at nIR
wavelengths is roughly a factor of ten lower than at optical
wavelengths, but it is smaller than a factor of two if the temperature
contrast is 1000\,K or higher.  Unfortunately, not much is known about
spot temperatures, particularly in low-mass stars, but our results
show that simultaneous observations at optical and nIR wavelengths can
provide useful constraints on the spot temperatures of active stars.

Another important factor for the effect of active regions on radial
velocity measurements are the differences between spectral features
appearing in the photosphere and the spots. Conventional estimates
usually assume that both are comparable, but given the perhaps
relatively large temperature contrasts and the strong temperature
dependence of the molecular features, this may not be the case. Thus,
large differences in the radial velocity signal between different
spectral regions can occur if spectral features of different
temperature sensitivity appear. 

The radial velocity signal may not vanish as expected at nIR
wavelengths, and it seems unlikely that strong radial velocity signals
observed at optical wavelengths can vanish in the nIR, particularly in
very active stars in which relatively large temperature contrast is
expected. The advantage of a nIR spectrograph over an optical
spectrograph becomes obvious in the late-M dwarfs. Our results point
towards a spectrograph covering the wavelength range 500--1150\,nm
that captures the region where the RV precision is highest at all M
spectral classes, and where the wavelength dependece of jitter shows
the largest gradient in order to distingiush between orbital motion
and spots. Such a spectrograph should be designed to be very stable
and could use a ThAr lamp for calibration. In the future, other
calibration strategies might become available \citep[e.g. a laser
frequency comb,][]{Steinmetz08}, but the ThAr method can in principle
provide the sensitivity required to detect Earth-mass planets around
low mass stars.


\acknowledgements

We thank Peter H. Hauschildt for promptly providing PHOENIX model
spectra.  A.R. acknowledges research funding from the DFG as an Emmy
Noether fellow, A.R. and A.S. received support from the DFG under RE
1664/4-1.  J.B. has received research funding from the European
Commission’s Seventh Framework Programme as an International Incoming
Fellow (PIFF-GA-2009-234866).





\begin{thebibliography}{}
\bibitem[Allard et al., 2001]{Allard01}Allard, F., Hauschildt, P.H.,
  Alexander, D.R., Tamanai, A., \& Schweitzer, A., 2001, \apj, 556,
  357
\bibitem[D'Amato et al., 2008]{DAmato08}D'Amato, F., et al., 2008,
  SPIE, 7014, 70143V
\bibitem[Bean et al., 2009]{Bean09}Bean, J., Seifahrt, A., Hartmann,
  H., Nilsson, H., Wiedemann, G., Reiners, A., Dreizler, S., \& Henry,
  T., submitted to ApJ, \texttt{arXiv:0911.3148}
\bibitem[Bouchy et al., 2001]{Bouchy01}Bouchy, F., Pepe, F., \&
  Queloz, D., 2001, \aap, 374, 733
\bibitem[Butler et al., 1996]{Butler96}Butler, R.P., Marcy, G.W.,
  Williams, E., McCarthy, C., Dosanjh, P., \& Vogt, S.S., 1996, PASP,
  108, 500
\bibitem[Connes, 1985]{Connes85}Connes, P., 1985, Ap\&SS, 110 211
\bibitem[Clough et al.(1981)]{Clough81} Clough, S.~A., Kneizys,
F.~X., Rothman, L.~S., \& Gallery, W.~O.\ 1981, \procspie, 277, 152
\bibitem[Clough et al.(1992)]{Clough92} Clough, S.~A., Iacono,
M.~J., \& Moncet, J.-L.\ 1992, \jgr, 97, 15761
\bibitem[Cushing et al., 2005]{Cushing05}Cushing, M.C., Rayner, J.T.,
  \& Vacca, W.D., 2005, \apj, 623, 1115
\bibitem[Delfosse et al., 1998]{Delfosse98}Delfosse, X., Forveille,
  T., Perrier, C., \& Mayor, M., 1998, \aap, 331, 581
\bibitem[Desort et al., 2007]{Desort07}Desort, M., Lagrange, A.-M.,
  Galland, F., Udry, S., \& Mayor, M., 2007, A\&A, 473, 983
\bibitem[Donati et al., 2008]{Donati08}Donati, J.-F., Morin, J.,
  Petit, P., et al., 2008, MNRAS, 390, 545
\bibitem[Hauschildt et al., 1999]{Hauschildt99}Hauschildt, P.H.,
  Allard, F., \& Baron, E., 1999, \apj, 512, 377
\bibitem[Hinkle et al., 2003]{Hinkle03}Hinkle, K.H., Wallace, L.,
  Valenti, J., \& Tsuji, T., 2003, IAU Symp. 215, 213
\bibitem[K\"{a}ufl et al., 2006]{Kaufl06}K{\"{a}ufl}, H.~U., et al.,
  2006, Msngr, 126, 32
\bibitem[Kerber et al., 2008]{Kerber08}Kerber, F., Nave, G., \&
  Sansonetti, C. J., 2008, \apjs, 178, 374
\bibitem[Lovis et al., 2006]{Lovis06}, Lovis., C., Pepe, F., Bouchy,
  F., et al., 2006, Proc. SPIE, Vol. 6269, 62690P
\bibitem[Lovis \& Pepe, 2007]{Lovis07}Lovis, C., \& Pepe, F., 2007,
  \aap, 468, 1115
\bibitem[Mahadevan \& Ge, 2009]{Mahadevan09}Mahadevan, S., \& and Ge,
  J., 2009, \apj, 692, 1590
\bibitem[Maltby et al., 1986]{Maltby86}Maltby, P., Avrett, E.H.,
  Carlsson, M., Kjeldseth-Moe, O., Kurucz, R.L., \& Loeser, R., 1986,
  \apj, 306, 284
\bibitem[Marcy et al., 1998]{Marcy98}Marcy, G.W., Butler, R.P., Vogt,
  S.S., Fischer, D., Lissauer, J.J., 1998, \apj, 505, L147
\bibitem[Mart\'in et al., 2006]{Martin06}Mart\'in, E.L., Guenther, E.,
  Zapatero Osorio, M.R., Bouy, \& Wainscoat, R., 2006, \apj, 644, L75
\bibitem[Mayor \& Queloz, 1995]{Mayor95}Mayor, M., \& Queloz, D.,
  1995, Nature, 378, 355
\bibitem[Mayor et al., 2003]{Mayor03}Mayor, M., et al., 2003, Msngr,
  124, 20
\bibitem[McLean et al., 2003]{McLean03}McLean, I.S., McGovern, M.R.,
  Burgasser, A.J., Kirkpatrick, J.D., Prato, L., \& Kim, S.S., 2003,
  \apj, 596, 561
\bibitem[McLean et al., 2007]{McLean07}McLean, I.S., Prato, L.,
  McGovern, M.R., Burgasser, A.J., Kirkpatrick, J.D., Rice, E.L., \&
  Kim, S.S., 2007, \apj, 658, 1217
\bibitem[Mohanty \& Basri, 2003]{Mohanty03}Mohanty, S., \& Basri, G.,
  2003, \apj, 583, 451
\bibitem[O'Neal et al., 2001]{ONeal01}O'Neal, D., Neff, J.E., Saar,
  S.H., \& Mines, J.K., 2001, \aj, 122, 1954
\bibitem[O'Neal et al., 2004]{ONeal04}O'Neal, D., Neff, J.E., Saar,
  S.H., \& Cuntz, M., 2004, \aj, 128, 1802
\bibitem[Pepe et al., 2002]{Pepe02}Pepe, F., Mayor, M., Galland, F.,
  Naef, D., Queloz, D., Santos, N.C., Udry, S., \& Burnet, M., 2002,
  A\&A, 388, 632
\bibitem[Piskunov \& Rice, 1993]{Piskunov93}Piskunov, N.E., \& Rice,
  J.B., 1993, PASP, 105, 1415
\bibitem[Reiners, 2009]{Reiners09a}Reiners, A., 2009, \aap, 498, 853
\bibitem[Reiners \& Basri, 2006]{Reiners06}Reiners, A., \& Basri, G.,
  2006, \apj, 644, 497
\bibitem[Reiners \& Basri, 2008]{Reiners08}Reiners, A., \& Basri, G.,
  2008, \apj, 684, 1390
\bibitem[Reiners \& Basri, 2009]{RB09}Reiners, A., \& Basri, G., 2009,
  A\&A, 496, 787
\bibitem[Reiners \& Basri, submitted to ApJ]{Reiners09}Reiners, A., \&
  Basri, G., submitted to ApJ
\bibitem[Rothman et al.(2005)]{HITRAN2004} Rothman, L.~S., et al.\
2005, Journal of Quantitative Spectroscopy and Radiative Transfer, 96, 139
\bibitem[Seifahrt et al., submitted to A\&A]{Seifahrt10}Seifahrt, A.,
  K\"aufl, H.U., Bean J., Richter, \& Siebenmorgen, R., 2010,
  submitted to A\&A
\bibitem[Solanki, 2003]{Solanki03}Solanki, S.K., 2003, A\&AR, 11, 153
\bibitem[Steinmetz et al.(2008)]{Steinmetz08} Steinmetz, T., Wilken,
  T., Araujo-Hauck, C., Holzwarth, R., H\"{a}nsch, T. W., Pasquini,
  L., Manescau, A., \& et al. 2008, Science, 321, 1335
\bibitem[Strassmeier \& Rice, 1998]{Strassmeier98}Strassmeier, K.G.,
  \& Rice, J.B., 1998, \aap, 330, 685
\bibitem[Tarter et al., 2007]{Tarter07}Tarter, J.C., et al., 2007,
  Astrobiology, 7, 30
\bibitem[Udry et al., 2007]{Udry07}Udry, S., Bonfils, X., Delfosse,
  X., et al., 2007, \aap, 469, L43
\bibitem[Wallace \& Livingston, 1992]{Wallace92}Wallace, L., \&
  Livingston, W., 1992, N.S.O.  Technical Report \#92-001
\bibitem[Wallace \& Hinkle, 1996]{Wallace96}Wallace, L., \& Hinkle,
  K.H., 1996, \apjs, 107, 312
\bibitem[Wallace et al., 1998]{Wallace98}Wallace, L., Livingston, W.,
  Bernath, P.F., \& Ram, R.S., 1998, N.S.O.  Technical Report
  \#1998-002
\bibitem[Zapatero-Osorio et al., 2007]{Zapatero07}Zapatero-Osorio
  M.R., Mart\'in, E.L., B\'ejar, V.J., Bouy, H., Deshpande, R., \&
  Wainscoat, R.J., 2007, \apj, 666, 1205
\end{thebibliography}
\end{document}